\documentclass[prb,aps,twocolumn,showpacs,amsmath,floatfix]{revtex4}
\usepackage{bbm}
\usepackage{amssymb}
\usepackage{amstext}
\usepackage{amscd}
\usepackage{amsthm}
\usepackage{amsopn}
\usepackage{amsmath}
\usepackage{amsbsy}
\usepackage{indentfirst}
\usepackage{graphicx}
\usepackage{verbatim}
\usepackage{bm}
\usepackage[bookmarks=false]{hyperref}

\begin{document}
\author{Sung-Po Chao, Vivek Aji}
\affiliation{Department of Physics and
Astronomy, University of California, Riverside, CA 92521}
 \title{Kondo and charge fluctuation resistivity due to Anderson impurities in graphene}

\begin{abstract}
Motivated by experiments on ion irradiated graphene, we compute the resistivity of  graphene with dilute impurities. In the local moment regime we employ the perturbation theory up to third order in the exchange coupling to determine the behavior at high temperatures within the Kondo model. Resistivity due to charge fluctuations is obtained within the mean field approach on the Anderson impurity model. Due to the linear spectrum of the graphene the Kondo behavior is shown to depend on the gate voltage applied. The location of the impurity on the graphene sheet is an important variable determining its effect on the Kondo scale and resitivity. Our results show that for chemical potential near the node the charge fluctuations is responsible for the observed temperature dependence of resistivity while away from the node the spin fluctuations take over. Quantitative agreement with experimental data is achieved if the energy of the impurity level varies linearly with the chemical potential.
\end{abstract}
\maketitle
\section{Introduction}
A logarithmic upturn in the resistivity at low temperature has been observed in graphene with vacancies\cite{Chen}. A fit to the temperature dependence of resistivity with conventional Kondo effect yields a large Kondo temperature (with $T_k\simeq 30\sim 90 K$) which shows a non-monotonic behavior with respect to the gate voltage\cite{Chen}. The vacancies in the graphene sheets are induced by ion irradiation in ultra-high vacuum and the magnetism in sputtered graphite 
has been experimentally observed\cite{Lehtinen,Oleg,PE,Ugeda,JJ}. Our goal is to study whether Kondo effect alone in graphene can explain the experimental results in Ref.\onlinecite{Chen}. 

We start with the Anderson impurity model\cite{Bruno,Bruno2} to study the impurities effect on transport. In the local moment regime where the impurity occupation for a given spin $n_{d,s}\simeq 0.5$ we use Schrieffer Wolff transformation\cite{Sch} to write down the Kondo model from the Anderson impurity Hamiltonian. Since we are interested in the resistivity due to impurity spin fluctuations we study the Kondo model by standard perturbation method. The perturbative approach fails at $T\simeq T_k$ with $T_k$ representing Kondo temperature but works for $T_k\ll T$. We compute 
the scattering rates in this weak coupling regime. The scattering rates are determined via perturbative calculations of the $T$-matrix\cite{Fischer,Nagaoka,Suhl}.
Kondo effect in the pseudogap system has been explored in the context of graphene as well as in that of d-wave superconductor\cite{DW,Ingersent,Carlos,Anatoli,Sengupta,Vojta,Vojta2,Fritz} via various different approach such as NRG or mean field approach\cite{Bruno,Bruno2}. The advantage of our   approach is the ability to determine the high temperature behavior of the scattering rate and resistivity accurately within perturbation theory.

We assume a dilute concentration of impurities and ignore the spin spin interactions such as the Ruderman-Kittel-Kasuya-Yosida (RKKY) interaction. In graphene these interactions, in addition to being oscillatory with distance between impurities, depend on the sublattice on which the impurities are located\cite{Brey,Saremi}.
For chemical potential at the Dirac point our results are in agreement with the prediction of the existence of an intermediate coupling fixed point\cite{DW,Ingersent,Carlos,Anatoli}. Near the node the exchange coupling $J$ needs to be larger than a critical value $J_c$ to have the Kondo effect. The dependence of $T_k$ on the chemical potential is qualitatively different for $\mu\ll T_k$ and $\mu\gg T_k$. For impurities breaking the lattice symmetry, a power law in $T$ divergence of scattering rate is obtained for $\mu\ll T_k$ while a logarithmic divergence appears for $\mu\gg T_k$. For impurities preserving the lattice symmetry, a power law in $T^3$ divergence of scattering rate is obtained for $\mu\ll T_k$ while a similar logarithmic divergence appears for $\mu\gg T_k$. For both cases the scaling of resistivity with single Kondo temperature breaks down in the vicinity of the Dirac point. Our results for Kondo temperature obtained within the $T$-matrix formalism is in agreement with the mean field results for the development of the Kondo phase\cite{Bruno,Sengupta}. The resistivity obtained displays different chemical potential dependences. For impurities breaking the lattice symmetry the resistivity decreases as chemical potential increases while for impurities preserving the lattice symmetry the resistivity increases as chemical potential increases. For the same set of physical parameters the dominant source of resistivity is from impurities which breaks the lattice symmetry.

We also explore the region near the empty orbital to mixed valence one in the Anderson impurity model to find the resistivity due to charge fluctuations. From the numerical RG\cite{Ingersent,Carlos} the Kondo effect is suppressed as the critical exchange coupling $J_c\rightarrow \infty$ for chemical potential close to the Dirac point. Thus we
use unrestricted Hatree Fock method\cite{Anderson} on the Anderson impurity model to find the resistivity near the empty orbital regime. The resulting resistivity shows similar dependence on chemical potential as well as dominance from symmetry breaking impurities as the resistivity obtained in the Kondo model.
Near the node the Kondo scale, extracted from the logarithmic temperature dependence region on resistivity, yields a Kondo temperature comparable to the observations in the experiment in Ref.\onlinecite{Chen} while away from the Dirac point the extracted Kondo scale is higher than experiment by one order of magnitude.

By combining the charge fluctuation effect for $\mu\simeq 0$ and Kondo effect (spin fluctuations) for finite $\mu$ we obtain Kondo temperature dependence on $\mu$
qualitatively consistent with experimental results\cite{Chen} with gate voltage less than $30V$. Our conclusion is that the observed experimental results, albeit fitted well by Numerical RG for conventional metal Kondo model\cite{Costi}, cannot be solely explained by Kondo screening in all range of chemical potential. For chemical potential near the node the charge fluctuations is responsible for the observed resistivity temperature dependence while away from the node the spin fluctuations take over.

This article is organized as following: We start with the Anderson impurity Hamiltonian to describe dilute impurities physics in the graphene system. To study the local moment regime we use Schrieffer Wolff transformation to obtain Kondo model from Anderson Hamiltonian. In section $3$ we evaluate resistivity due to spin fluctuations, with different impurity locations, by perturbation computations on the Kondo model. In section $4$ we compute resistivity due to charge fluctuations when impurity occupation is close to zero by using mean field approach on the Anderson model. In section $5$ we show numerical results of temperature dependence of the resistivity with different symmetry and mechanism. In section $6$ we compare our results with the experiment in Ref.\onlinecite{Chen}.
The results are summarized in section $7$. Two appendixes contain derivations for the perturbative results in the Kondo model. 
\section{Hamiltonian}
We start from graphene Hamiltonian in the presence of dilute impurities described by the Anderson impurity Hamiltonian\cite{Bruno}
\begin{eqnarray}
&&H=H_g+H_d+H_U+H_V\\
&&H_g=-t\sum_{k,s}\phi_k a_{k,s}^{\dagger}b_{k,s}+h.c.-\mu(a_{k,s}^{\dagger}a_{k,s}+b_{k,s}^{\dagger}b_{k,s})\nonumber\\
&&H_d=\sum_s \epsilon_d d_s^{\dagger}d_s\nonumber\\
&&H_U=U d_s^{\dagger}d_s d_{-s}^{\dagger}d_{-s}\nonumber\\
&&H_V=\sum_{k,s}[V_{a,k}^{\ast}a_{k,s}^{\dagger}+V_{b,k}b_{k,s}^{\dagger}]d_s+h.c.\nonumber
\end{eqnarray}
$H_g$ is the nearest hopping in the momentum space with $t\simeq 2.7eV$ being the nearest neighbor hopping strength. $\mu$ defines the Fermi level measured from the Dirac point. $a^{\dagger}_{k,s}$ and $b^{\dagger}_{k,s}$ are the particle creation operators on the a and b sublattices. $\phi_k=\sum_{i=1}^3e^{i\vec{k}\cdot\vec{a}_i}$ with $\vec{a}_1=a_0\vec{x}$, $\vec{a}_2=a_0(-\vec{x}/2+\sqrt{3}\vec{y}/2)$, and $\vec{a}_3=a_0(-\vec{x}/2-\sqrt{3}\vec{y}/2)$ being the nearest neighbor lattice vector. $a_0\simeq 
1.42\dot A$ is the lattice constant. $H_V$ describes the hybridization between the impurity level and graphene electrons with $V_{k,a/b}=\sum_{i=1}^3 V_{i,a/b}e^{i\vec{k}\cdot\vec{a}_i}$. $H_U$ describes the Coulomb repulsion
on the impurity level and $H_d$ is the Hamiltonian describing the localized level of $d$ electron. We diagonalize $H_g$ by defining $c^\dagger_{ks\pm}=(a^\dagger_{ks}\pm(\phi_k/|\phi_k|)b^\dagger_{ks})$. In this basis the $H_g$ term becomes
\begin{eqnarray}
H_g=\sum_{k s n} (nt|\phi_k|-\mu)c_{ksn}^{\dagger}c_{ksn}
\end{eqnarray}
 with $n=\pm$ denoting the conduction and valence bands. The hybridization term $H_V$ in this rotated basis is
\begin{eqnarray*}
H_V=V\sum_{n=\pm}\sum_{k,s}[\Theta_{kn}c^{\dagger}_{ksn}d_s+h.c.]
\end{eqnarray*}
with $\Theta_{kn}=(V_{k,b}+nV_{k,a}^{\ast}\phi_k^{\ast}/|\phi_k|)/(\sqrt{2}V)$. Denote $\epsilon_{k,n}=nt|\phi_k|-\mu$ as the energy of the bands evaluated from chemical potential $\mu$ and $\{k\}=(\vec{k}sn)$ as combinations of momentum, spin, and $n$ the band index. The Anderson impurity Hamiltonian describing the impurity in the graphene can be written as 
\begin{eqnarray}\nonumber
H&=&\sum_{\{k\}}\epsilon_{kn}c_{k s n}^{\dagger}c_{k s n}+\sum_s\epsilon_d d_s^{\dagger}d_s+U d^{\dagger}_sd_sd^{\dagger}_{-s}d_{-s}\\\label{Anderson}
&+&V\sum_{\{k\}}[\Theta_{kn}c^{\dagger}_{ksn}d_s+\Theta_{kn}^{\ast}d_s^{\dagger}c_{ksn}]
\end{eqnarray}
To explore the local moment regime where impurity occupation for a given spin $n_{d,s}\simeq 0.5$ we perform Schrieffer Wolf transformation to project out the
charge degree of freedom\cite{Bruno}. 
The exchange Hamiltonian or Kondo model obtained after this transformation with the additional term $H_{int}\propto \vec{S}_r\cdot \vec{S}_{r'}$ describing 
spin spin interaction at different sites is given by
$H=H_g+H_{imp}+H_{int}$ with
\begin{eqnarray}\label{Kondo}
H_g&=&\sum_{\{k\}}\epsilon_{kn}c_{k s n}^{\dagger}c_{k s n}\\\nonumber
H_{imp}&=&\frac{1}{N}\sum_{\{k\},\{k'\}}\Theta_{kn}^{\ast}\Theta_{k'm}(K\delta_{s's}-\frac{J}{2}\vec{S}\cdot \vec{\sigma}_{s's} )c^{\dagger}_{k's' m}c_{ks n}\\\nonumber
H_{int}&=&-\sum_{r,r'}W(r-r')\vec{S}_r\cdot \vec{S}_{r'}
\end{eqnarray}
 Here $J \simeq V^2(1/(\epsilon_d-\mu)-1/(\epsilon_d+U-\mu))$ and $K \simeq (V^2/2)(1/(\mu-\epsilon_d)+1/(\mu-\epsilon_d-U))$.
  The interaction between impurity spins $H_{int}$ are added for the inclusion of spin spin interaction but is assumed to be small due to small concentration of the impurities in this article. Including this term would lead to a time dependent impurity spin via $\mathrm{S}(\tau)=e^{H_{int}\tau}\mathrm{S}e^{-H_{int}\tau}$ with $\tau$ being the imaginary time\cite{Fischer}.

For impurities preserving the $C_{3v}$ point group symmetry of the triangular sublattice in the graphene system the factor $|\Theta_{kn}|\propto|\phi_k|$ while for impurities breaking the symmetry the factor $|\Theta_{kn}|$ is a constant. To evaluate the resistivity due to spin fluctuations we use Eq.(\ref{Kondo}) as the starting Hamiltonian. We use perturbation expansion on the one particle Green function's T-matrix to compute scattering rate and from Boltzmann transport to obtain linear response resistivity in both impurity breaking and preserving the lattice symmetry cases. We use mean field approach on the Anderson impurity model shown in Eq.(\ref{Anderson}) to obtain resistivity due to charge fluctuations in both symmetry breaking and preserving case. The following two sections are 
the computation results for each cases mentioned above.

\section{Resistivity due to impurity spin fluctuations}
To study the resistivity due to spin fluctuations we start with the Kondo Hamiltonian shown in Eq.(\ref{Kondo}). 
We calculate transport properties from the $T$-matrix which is related to single particle Green's function by\cite{Fischer}
\begin{eqnarray}
&&G_{\{k'\},\{k\},\alpha',\alpha}(i\omega_1,i\omega_2)=G_{\{k'\},\{k\}}^0(i\omega_1)\\\nonumber&&+G_{\{k'\},\{k'\}}^0(i\omega_1)T_{\{k'\},\{k\},\alpha',\alpha}(i\omega_1,i\omega_2)
G_{\{k\},\{k\}}^0(i\omega_2)
\end{eqnarray}
where $G_{\{k'\},\{k\}}^0(i\omega_1)=\delta_{k',k}\delta_{s',s}\delta_{m,n}(i\omega_1-\epsilon_{k,n})^{-1}$ and $\alpha$, $\alpha'$ are impurity spin state indices. This expression is related to time dependent Green's function by
\begin{eqnarray}
&&G_{\{k'\},\{k\},\alpha',\alpha}(i\omega_1,i\omega_2)\\\nonumber&&=\int_0^\beta \int_0^\beta  d\tau d\tau' e^{i(\omega_1\tau-\omega_2\tau')}G_{\{k'\},\{k\},\alpha',\alpha}(\tau,\tau')
\end{eqnarray}
with $\omega=(2r+1)\pi T=(2r+1)\pi/\beta$ and $r$ being integers (We put Boltzmann constant $k_B=1$ to simplify the notation). Based on perturbation in $H_{int}$ this time dependent Green's function can be written as
\begin{widetext}
\begin{eqnarray}
&&G_{\{k'\},\{k\},\alpha',\alpha}(\tau,\tau')=-\langle Z\rangle^{-1}\sum_{n=0}^{\infty}\frac{(-1)^n}{n!}\int_0^{\beta} d\tau_1\ldots \int_0^\beta d\tau_n
\sum_{q_1,q_1',..,q_m,q_n'}\sum_{s_1,s_1',..,s_n,s_n'}\sum_{n_1,m_1,..,n_m,m_n}\\\nonumber&&[T_S(\Theta_{k_1,n_1}^{\ast}\Theta_{k_1',m_1}(K\delta_{s_1',s_1}-\frac{J}{2}\vec{S}(\tau_1)\cdot\vec{\sigma}_{s_1',s_1})\ldots\Theta^{\ast}_{k_n,n_n}\Theta_{k_n',m_n}(K\delta_{s_n',s_n}-\frac{J}{2}\vec{S}(\tau_n)\cdot\vec{\sigma}_{s_n',s_n}))]_{\alpha',\alpha}\\\nonumber
&&\langle T_{\tau}(c_{k' s' m}(\tau)\bar{c}_{k s n}(\tau')\bar{c}_{k_1' s_1' m_1}(\tau_1)c_{k_1 s_1 n_1}(\tau_1)\ldots
\bar{c}_{k_n's_n'm_n}(\tau_n)c_{k_n s_n n_n}(\tau_n)\rangle_{H_e}
\end{eqnarray}
\end{widetext}
Here $T_S$ and $T_{\tau}$ are the time ordering operators and $\langle Z\rangle$ is the S-matrix. The first order in $J$ and $K$ is given by
\begin{eqnarray}
T^{(1)}_{\{k'\},\{k\},\alpha',\alpha}(i\omega_1,i\omega_2)&=&\Theta_{kn}^{\ast}\Theta_{k'm}[\beta\delta_{\omega_1,\omega_2}K\delta_{s',s}\delta_{\alpha',\alpha}\nonumber\\&-&\frac{J}{2}\sigma_{s's}S_{\alpha'\alpha}(i(\omega_1-\omega_2))]
\end{eqnarray}
 Here $S(i\omega')=\int_0^\beta d\tau e^{i\omega'\tau}S(\tau)$ with $\omega'=2\pi r k_B T$. For $H_{RKKY}=0$ we can simplify above expression by noting that
$S(i\omega')=\beta S\delta_{\omega',0}$ and we get
\begin{eqnarray}
T^{(1)}_{\{k'\},\{k\},\alpha',\alpha}(i\omega_1,i\omega_2)=\beta\delta_{\omega_1,\omega_2}T^{(1)}_{\{k'\},\{k\},\alpha',\alpha}(i\omega_1)
\end{eqnarray}
with $T^{(1)}_{\{k'\},\{k\},\alpha',\alpha}(i\omega_1)=\Theta_{kn}^{\ast}\Theta_{k'm}(K-\frac{J}{2}\sigma_{s's}S_{\alpha'\alpha})$. The general second order result
of T-matrix, with $f_{k,n}\equiv 1/(e^{\beta\epsilon_{k,n}}+1)$ and $F(z)\equiv N^{-1}\sum_{k,n}|\Theta_{kn}|^2(z-\epsilon_{k,n})^{-1}$, is expressed as
\begin{widetext} 
\begin{eqnarray}\label{2ndop}
&&T^{(2)}_{\{k'\},\{k\},\alpha',\alpha}(i\omega_1,i\omega_2)=\Theta_{kn}^{\ast}\Theta_{k'm}\{\beta\delta_{\omega_1,\omega_2}\delta_{\alpha',\alpha}K^2-K J \sigma_{s's}S_{\alpha'\alpha}(i\omega_1-i\omega_2)\}F(i\omega_1)\\\nonumber&&+\Theta_{kn}^{\ast}\Theta_{k'm}\Big(\frac{J}{2}\Big)^2 T\sum_{\omega_1'\omega_2'}(\sigma^{i_1}\sigma^{i_2})_{s's}\delta_{\omega_1'+\omega_2',\omega_1-\omega_2}\frac{1}{N}\sum_{q,l}|\Theta_{ql}|^2 G^0_{q,l}(i\omega_1-i\omega_1')\{S^{i_1}(i\omega_1')S^{i_2}(i\omega_2')-f_{q,l}[S^{i_1}(i\omega_1'),S^{i_2}(i\omega_2')]\}_{\alpha'\alpha}
\end{eqnarray}
\end{widetext}
To focus on the Kondo contribution to the scattering rate we may set $K=0$ and we set $H_{int}=0$ by assuming dilute impurities. For RKKY type of spin spin interactions the interaction strength decays as $1/R^3$ for symmetry breaking or $1/R^7$ for symmetry preserving case\cite{Bruno}. Thus for sufficient dilute impurities we may treat $H_{int}=0$. In this limit Eq.(\ref{2ndop}) is simplified to $T^{(2)}_{\{k'\},\{k\},\alpha',\alpha}(i\omega_1,i\omega_2)=\beta\delta_{\omega_1,\omega_2} T^{(2)}_{\{k'\},\{k\},\alpha',\alpha}(i\omega_1)$
with 
\begin{eqnarray}
&&T^{(2)}_{\{k'\},\{k\},\alpha',\alpha}(z)=\Theta_{kn}^{\ast}\Theta_{k'm}\Big(\frac{J}{2}\Big)^2[S(S+1)F(z)\delta_{s',s}\delta_{\alpha',\alpha}\nonumber \\\label{2nd}&&+\frac{1}{N}\sum_{q,l}\frac{|\Theta_{ql}|^2}{\epsilon_{q,l}-z}\tanh(\frac{\beta\epsilon_{q,l}}{2})(S_{\alpha'\alpha}\cdot\sigma_{s's})]
\end{eqnarray}
For noninteracting spins the third order perturbation, after taking a trace over conduction electron spins and approximating the reduction of three-spin correlation functions to two-spin correlation functions\cite{Fischer}, is given by
\begin{eqnarray}
&&T^{(3)}(z)=2S(S+1)\Big(\frac{J}{2}\Big)^3\Theta_{kn}^{\ast}\Theta_{k'm}\frac{1}{N^2}\sum_{q_1,n_1}\Big[\frac{|\Theta_{q_1n_1}|^2}{z-\epsilon_{q_1,n_1}}
\nonumber\\\label{3rd}&&\sum_{q_2,n_2}\frac{|\Theta_{q_2n_2}|^2\tanh(\frac{\beta\epsilon_{q_2,n_2}}{2})}{\epsilon_{q_1,n_1}-\epsilon_{q_2,n_2}}\Big]
\end{eqnarray}
From Eq.(\ref{2nd}) and Eq.(\ref{3rd}) we may define a general function $R(z)$ which we need to evaluate in computing the $T$-matrix
\begin{eqnarray}\label{expR}
R(z)=\frac{1}{N}\sum_{q,n}\frac{|\Theta_{qn}|^2}{\epsilon_{q,n}-z}\tanh\Big(\frac{\beta\epsilon_{q,n}}{2}\Big)
\end{eqnarray}
 By using $\epsilon_{q,n}=nt|\phi_q|-\mu\simeq n\frac{3ta_0}{2}|q|-\mu=n|\epsilon|-\mu$ we may write the continuous form of Eq.(\ref{expR})
as
\begin{eqnarray}\label{expRc}
R(z)=\frac{8}{9\pi t^2}\int_{-\Lambda}^{\Lambda} d\epsilon |\epsilon|\frac{|\Theta_{\epsilon}|^2}{\epsilon-(z+\mu)}\tanh\Big(\frac{\beta(\epsilon-\mu)}{2}\Big)
\end{eqnarray}
where $\Lambda$ is the linear spectrum cutoff.
 In the following we will separate the discussions into two cases\cite{Bruno}: One with impurity interactions breaking the $C_{3v}$ lattice symmetry, in which $|\Theta_{kn}|^2=\frac{1}{2}$, and we denote $R(z)=R_{SB}(z)$ in this case. Another with impurity interactions preserving the symmetry, in which $|\Theta_{kn}|^2=\frac{9|k|^2a_0^2}{8}=\frac{|\epsilon|^2}{2t^2}$, and we denote $R(z)=R_{SP}(z)$ in this case.

\subsection{$C_{3v}$ symmetry breaking impurities}
For the case of symmetry breaking the cutoff scheme we choose for a linear density of states with a cutoff $\Lambda$ is multiplying the argument of right hand side of Eq.(\ref{expRc}) by $\frac{\Lambda^2}{\epsilon^2+\Lambda^2}$ and extend the integration limit from $\pm\Lambda$ to $\pm\infty$. The resulting $R_{SB}(z)$, with details shown in
Appendix. \ref{A}, is 
\begin{eqnarray}
&&R_{SB}(z)=\frac{4}{9\pi t^2}\Big\{\frac{\Lambda}{\pi}\Big[P\int_0^{\infty}\frac{x dx}{x-\Lambda}F(x,\mu,z)\\\nonumber&&-\int_0^{\infty}\frac{x dx}{x+\Lambda}F(x,\mu,z)\Big]-\psi(\frac{1}{2}-i\frac{\beta z}{2\pi})\frac{2\Lambda^2(z+\mu)}{(z+\mu)^2+\Lambda^2}\Big\}
\end{eqnarray}
Here $F(x,\mu,z)$ is defined as
$$F(x,\mu,z)=\frac{\psi(\frac{1}{2}+i\frac{\beta\mu}{2\pi}+\frac{\beta x}{2\pi})}{x+i(\mu+z)}+
\frac{\psi(\frac{1}{2}-i\frac{\beta\mu}{2\pi}+\frac{\beta x}{2\pi})}{x-i(\mu+z)}$$
and $\psi(z)$ is the digamma function. Analytic forms can be obtained in two asymptotic limits by using the asymptotic forms of the digamma function.
For $\beta\mu\ll 2\pi$ we have
\begin{eqnarray}
&&R_{SB}(z)\simeq\frac{4}{9\pi t^2}\Big[\pi \Lambda+\frac{4\gamma+4\ln(2)-4-3\zeta(2)}{\beta}\nonumber\\\nonumber  
&&-\frac{(4\gamma-2\zeta(2)+8)(z+\mu)}{\pi}\tan^{-1}(\frac{\pi}{\beta(z+\mu)})\\\nonumber 
&&+\frac{\beta z(z+\mu)}{2}\ln(1+(\frac{\pi}{\beta(z+\mu)})^2)\\\label{bml}
&&-\frac{2\Lambda^2(z+\mu)}{(z+\mu)^2+\Lambda^2}\psi(\frac{1}{2}-i\frac{\beta z}{2\pi})\Big]
\end{eqnarray} 
where $\gamma\simeq 0.577$ is the Euler constant and $\zeta(2)=\pi^2/6$ is the Riemann zeta function evaluated at 2. In the limit $\beta\mu\gg 2\pi$ we get
\begin{eqnarray}\nonumber 
&&R_{SB}(z)\simeq\frac{4}{9\pi t^2}\Big[\frac{4}{\pi}\ln(\frac{\beta|\mu|}{2\pi})(z+\mu)\tan^{-1}(\frac{|\mu|}{z+\mu})\\\nonumber 
&&+\pi\Lambda+\frac{4|\mu|}{\pi}(\frac{\pi}{2}-\tan^{-1}(\frac{|\mu|}{z+\mu})-1)\\\nonumber 
&&-|z+\mu|\ln(1+\frac{\mu^2}{\mu^2+z^2})\\\label{bmg}
&&-\frac{2\Lambda^2(z+\mu)}{(z+\mu)^2+\Lambda^2}\psi(\frac{1}{2}-i\frac{\beta z}{2\pi})\Big]
\end{eqnarray}
In Eq.(\ref{bml}) and Eq.(\ref{bmg}) we have assumed $0\le z\le\mu$. 
Using the Boltzmann equation with relaxation time hypothesis\cite{HewsonBook} and noticing that the honeycomb symmetry is broken by the impurity we find the scattering rate is related to the $T$-matrix by 
\begin{eqnarray}\label{reltimesb}\nonumber
&&\frac{1}{\tau_{SB}(\epsilon_{k,n})}\\\nonumber
&&=\frac{\pi n_I}{\hbar}\int \delta(\epsilon_{k,n}-\epsilon_{k',m})|T_{k,k'}|^2(1-\cos\theta_{k,k'})\frac{d\vec{k'}}{(2\pi)^2}\\\nonumber
&&+\frac{\pi n_I}{\hbar}\int \delta(\epsilon_{k,n}-\epsilon_{k',m})|T_{k,k'}|^2(1-\cos\theta_{k,k'})\frac{d\vec{k'}}{(2\pi)^2}\\
&&=-\frac{2 n_I}{\hbar}\Im[T_{k,k}(\epsilon_{k,n})]
\end{eqnarray}
The second line of Eq.(\ref{reltimesb}) represents the scattering process related to different Dirac points in the Brillouin zone and the third line is the
scattering event within the same Dirac cone. We have used the fact that $\Im[T_{k,k}(\epsilon^+)]=-\pi\sum_{k'}|T_{k,k'}(\epsilon^+)|^2\delta(\epsilon-\epsilon_{k'})$ and $|T_{k,k'}(\epsilon^+)|^2$ is independent of 
angle $\theta_{k,k'}$ between momenta $k$ and $k'$ in the symmetry breaking case in the above equation.
The scattering rate, $\tau_{SB}^{-1}(\omega)$ with $\tau_{SB}(\omega)$ being the relaxation time, to third order is 
\begin{eqnarray}\nonumber
\hbar\tau_{SB}^{-1}(\omega)&=&4 n_I S(S+1)\Big[\Big(\frac{2}{9 t^2}\Big)\Big(\frac{J}{2}\Big)^2|\omega+\mu|\\
&-&2\Big(\frac{2}{9 t^2}\Big)\Big(\frac{J}{2}\Big)^3|\omega+\mu|\Re[R(\omega)]\Big]
\end{eqnarray}
The Kondo effect is reflected in the divergence of the relaxation time in the parquet approximation. This involves treating the cubic term as the
first in an infinite series which is summed to give
\begin{eqnarray}\label{relsb}
\hbar\tau_{SB}^{-1}(\omega)=4 n_I S(S+1)\Big(\frac{2}{9 t^2}\Big)\Big(\frac{J}{2}\Big)^2\frac{|\omega+\mu|}{1+J\Re[R(\omega)]}
\end{eqnarray}
\begin{figure}[t]
\includegraphics[width=1\columnwidth, clip]{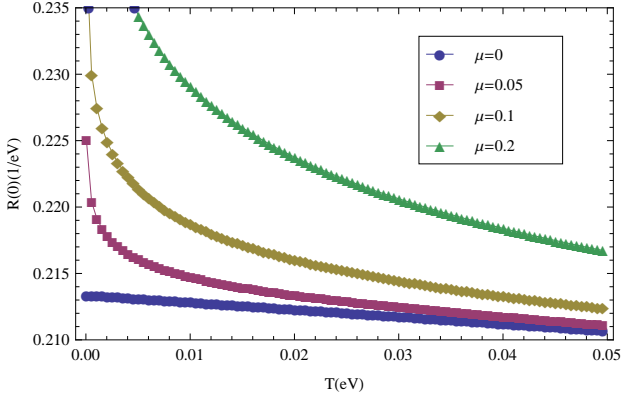}
\caption{The function $R_{SB}(0)$ plotted as a function of temperature $k_B T$ in the unit of $eV$. Energy cutoff $\Lambda=3.5eV$ and $t=2.7eV$.}
	  \label{fig1}
\end{figure}
We are mainly interested in the DC response so we shall study the relaxation time when $\omega\rightarrow 0$. Fig.\ref{fig1} shows the function $R_{SB}(0)$ plotted as a function of temperature for different chemical potential. We define the Kondo temperature as the temperature when the relaxation time diverges when $\omega\rightarrow 0$. For the case $\beta\mu\ll 2\pi$ the singularities from $1+J\Re[R(0)]\simeq 0$ can be expressed, by using Eq.(\ref{bml}), as
\begin{eqnarray}\nonumber
&&\frac{4J}{9\pi t^2}\Big[\pi\Lambda+\frac{4\gamma+4\ln(2)-4-3\zeta(2)}{\beta_k}\Big]\simeq -1\\\label{tksbs}
&&\rightarrow T_k=-\frac{\pi\Lambda}{4\gamma+4\ln(2)-4-3\zeta(2)}\Big(1-\frac{J_c}{J}\Big)
\end{eqnarray}
 where $J_c=-9t^2/4\Lambda$ and $\beta_k=1/T_k$. Thus for chemical potential $\mu\ll  T$ we have no Kondo effect if $|J|<|J_c|$. As one increases the chemical potential we may include the linear order of $\mu$ in Eq.(\ref{bml}) and obtain the expression for Kondo temperature as
\begin{eqnarray}\label{tksbm}
 T_k=-\frac{\pi\Lambda(1-\frac{J_c}{J})+(4\ln(2)+\zeta(2)-4)\mu}{4\gamma+4\ln(2)-4-3\zeta(2)}
\end{eqnarray}
where we have used $\tan^{-1}(\frac{\pi}{\beta(z+\mu)})\rightarrow \frac{\pi}{2}$ and $\psi(\frac{1}{2})=-\gamma-2\ln(2)$. Thus the Kondo temperature increases  with increasing chemical potential.

In the opposite limit where $\beta\mu\gg 2\pi$ we use Eq.(\ref{bmg}) to obtain the Kondo temperature as
\begin{eqnarray}\label{logtk}
 T_k=c_1\mu\exp\Big[\frac{\pi\Lambda}{\mu}\Big(1-\frac{J_c}{J}\Big)\Big]
\end{eqnarray}

\begin{figure}[t]
\includegraphics[width=1\columnwidth, clip]{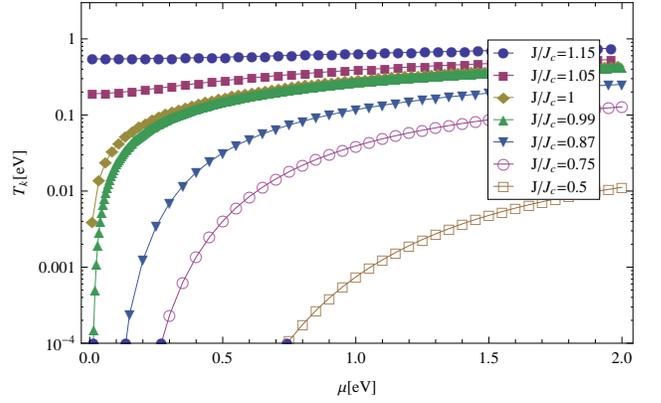}
\caption{The Kondo temperature as a function of chemical potential for various values of $J$. Energy cutoff $\Lambda=3.5eV$ and $t=2.7eV$ gives $|J_c|\simeq 4.68eV$.}
	  \label{fig2}
\end{figure}
\noindent where $c_1=\exp(\gamma+\ln(2)-1-4/\pi)/2\pi\simeq 0.058$.  Eq.(\ref{logtk}) can be expressed as $T_k\propto \exp((J-J_c)/(\rho_g J)$ with $\rho_g\propto\mu$ is the electron density of states of  graphene. Compared with the Kondo temperature in conventional metal $T_k\propto\exp(1/\rho_g J)$ there exists critical value of exchange coupling for Kondo effect to be realized in this two dimensional pseudo gap system. Fig.\ref{fig2} shows the Kondo temperature as a function of the chemical potential for various values 
of $J$. For $J/J_c$ smaller than $0.87$, the Kondo temperature is smaller than the chemical potential for the range shown. In this regime $T_k$ is given by
Eq.(\ref{logtk}) and is exponentially smaller than the energy scale set by the chemical potential. As $J/J_c$ approaches one from below, the Kondo temperature
grows faster than the chemical potential. As $T_k(\mu)\simeq \mu/2\pi$ the exponential behavior cross over to the linear dependence shown in Eq.(\ref{tksbm}).

Given the relaxation time we obtain the linear response conductivity as
\begin{eqnarray}\label{sbcond}
\sigma_{SB}^s(T)&=&-\frac{2e^2}{3}\sum_n\int v_F^2\tau_{SB}(\epsilon_{k,n})\frac{\partial f}{\partial \epsilon_{k,n}}\frac{d\vec{k}}{(2\pi)^2}\\\nonumber
&=&-\frac{2e^2}{3}\frac{4v_F^2}{9\pi a_0^2t^2}\int_{-\infty}^{\infty}d\epsilon|\epsilon|\tau_{SB}(\epsilon-\mu)\frac{\partial f}{\partial \epsilon}
\end{eqnarray}
For small chemical potential or $\beta\mu\ll 2\pi$ we use Eq.(\ref{bml}) and Eq.(\ref{tksbm}) and approximate $\int g(\epsilon)(-\partial f/\partial \epsilon)\simeq \int_{\mu-T}^{\mu+T} g(\epsilon)/(2T)$. We get the resistivity at low chemical potential as
\begin{eqnarray}\nonumber
\rho_{SB}^s(T)&\simeq&\frac{3}{2e^2}\frac{\pi S(S+1)a_0^2 n_I}{\hbar v_F^2}\left(\frac{J}{2}\right)^2\frac{9\pi t^2}{2J}\left(\frac{1}{r(T,\mu)}\right)\\\label{sbr}
&\simeq&\frac{3\pi n_I S(S+1)}{4}\left(\frac{J}{2}\right)\frac{h}{e^2}\left(\frac{1}{r(T,\mu)}\right)
\end{eqnarray}
with
\begin{eqnarray*}
r(T,\mu)&=&(4\gamma+4\ln(2)-4-3\zeta(2))(T-T_k)\\&-&(2\gamma-\zeta(2)+4-2\psi(\frac{1}{2}))\frac{(T-\mu)^2}{2T}
\end{eqnarray*}
 For temperature $\mu/2\pi< T<\mu$ but higher than $T_k$ the same approximation scheme gives
\begin{eqnarray}\nonumber
&&\rho_{SB}^s(T)\simeq\frac{3\pi n_I S(S+1)}{4}\left(\frac{J}{2}\right)\frac{h}{e^2}\left(\frac{1}{r(T,\mu)}\right).\\
&&r(T,\mu)=(4\gamma+4\ln(2)-4-3\zeta(2))(T-T_k)
\end{eqnarray}
Thus we see that for $T>\mu$ the Kondo contribution to resistance is not determined by a single scale $T_k$. For temperature ranged between $\mu/2\pi< T<\mu$ the scaling of the resistivity goes like $1/(T-T_k)$. This power law behavior indicates that at sufficient low chemical 
potential the magnetic impurities are not completely quenched while a logarithmic behavior is expected in the conventional metal case.

For large $\mu$ or $\beta\mu\gg 2\pi$ a Kondo effect similar to magnetic impurities in the conventional metals is obtained. For large chemical potential we approximate $\partial f/\partial \epsilon\simeq -\delta(\epsilon-\mu)$. Under this approximation the resistivity $\rho_{SB}^s(T)=\frac{1}{\sigma_{SB}^s(T)}$ is
given by
\begin{eqnarray}
\rho_{SB}^s(T)\simeq\frac{3}{2e^2}\frac{\pi S(S+1)a_0^2 n_I}{\hbar v_F^2}\frac{\left(\frac{J}{2}\right)^2}{1+J\Re(R_{SB}(0))}
\end{eqnarray}
Use Eq.(\ref{bmg}) for $\Re(R_{SB}(0))$ with $T>T_k$ we get
\begin{eqnarray}\nonumber
\rho_{SB}^s(T)&\simeq&\frac{3}{2e^2}\frac{\pi S(S+1)a_0^2 n_I}{\hbar v_F^2}\left(\frac{J}{2}\right)^2\frac{9\pi t^2}{2J\mu}\left(\ln\left[\frac{T_k}{T}\right]\right)^{-1}\\
&\simeq&\frac{3\pi n_I S(S+1)}{4}\left(\frac{J}{2\mu}\right)\frac{h}{e^2}\left(\ln\left[\frac{T_k}{T}\right]\right)^{-1}
\end{eqnarray}

\subsection{$C_{3v}$ symmetry preserving impurities}
For the case of impurities preserving the symmetry of honeycomb lattice the cutoff scheme we choose for a linear density of states with a cutoff $\Lambda$ is multiplying the argument of right hand side of Eq.(\ref{expRc}) by $\Lambda^4/(\epsilon^4+\Lambda^4)$ and extend the integration limit from $\pm\Lambda$ to $\pm\infty$. The resulting $R_{SP}(z)$, with details shown in
Appendix \ref{B}, is 
\begin{eqnarray}\nonumber
&&R_{SP}(z)=\frac{4}{9\pi t^4}\Big\{\frac{2\Lambda^4}{\pi}\int_0^{\infty}dx \frac{x^3}{x^4+\Lambda^4}F(x,\mu,z)\\&&-2\frac{(z+\mu)^3\Lambda^4}{\Lambda^4+(z+\mu)^4}
\psi(\frac{1}{2}-i\frac{\beta z}{2\pi})\\\nonumber&&+\Re[\frac{\Lambda^4\psi(\frac{1}{2}+i\frac{\beta\mu}{2\pi}-i\frac{\beta\Lambda}{2\pi}e^{i\frac{3\pi}{4}}}{\Lambda e^{i\frac{3\pi}{4}}-(z+\mu)}-\frac{\Lambda^4\psi(\frac{1}{2}-i\frac{\beta\mu}{2\pi}-i\frac{\beta\Lambda}{2\pi}e^{i\frac{\pi}{4}}}{\Lambda e^{i\frac{\pi}{4}}-(z+\mu)}]\Big\}
\end{eqnarray}
Analytic forms of $R_{SP}(z)$ is obtained by taking the asymptotic behavior of digamma function in the following two limits: $\beta\mu/2\pi\ll 1$ and  $\beta\mu/2\pi\gg 1$. For $\beta\mu/2\pi\ll 1$ we have
\begin{eqnarray}\nonumber
&&R_{SP}(z)\simeq
\frac{4 \Lambda ^3} {9 \pi  t^4}\Big\{\frac{\pi}{\sqrt{2}}+\frac{(\pi-4\ln(\frac{2\pi}{\beta\Lambda}))\frac{z\mu}{\Lambda^2}}{\sqrt{2}}\\\nonumber
&&+\left(1-\frac{\left(4+8\ln\left(\frac{z}{\Lambda}\right)\right)z}{\pi\Lambda}\right)\frac{\pi}{\beta\Lambda}+\frac{\sqrt{2}}{3}\left(1+\frac{2z\mu}{\Lambda^2}\right)\left(\frac{\pi}{\beta\Lambda}\right)^2\\\nonumber
&&+\left(\frac{4(1+2\ln(2))}{9\pi}+\frac{\mu}{2\pi z}\right)\left(\frac{\pi}{\beta\Lambda}\right)^3\\\label{bmml}&&+O((\frac{\mu}{\Lambda})^2,(\frac{z}{\Lambda})^2)\Big\}  
\end{eqnarray}
For the opposite limit $\beta\mu/2\pi\gg 1$ we have
\begin{eqnarray}\nonumber
&&R_{SP}(z)\simeq\frac{4\Lambda^3}{9\pi t^4}\Big\{\frac{\pi}{\sqrt{2}}+\frac{2\pi}{\beta\Lambda}+\frac{(z+\mu)}{\pi\Lambda}\Big(\frac{2\mu}{\Lambda}-(2+\pi)\frac{\mu^2}{\Lambda^2}
\\\nonumber&&+2(1+\frac{\mu^2}{\Lambda^2})\tan^{-1}(\frac{\mu}{\Lambda})\Big)+\frac{z\mu}{\sqrt{2}\Lambda^2}(\pi-2-4\ln(\frac{\pi}{\beta\Lambda}))
\\\nonumber&&+\frac{4}{\Lambda^3\pi}\Big(-\mu  (z+\mu )^2+(z+\mu )^3 \tan ^{-1}\left(\frac{\mu }{z+\mu }\right)\\\nonumber&&+\frac{\mu ^3}{3}\Big) \ln\left(\frac{\beta  \mu }{2 \pi
   }\right)+\frac{\mu^2(\mu+z)}{\Lambda^3}
\\\label{bmmg}&&
+\frac{(\mu+z)^3}{\Lambda^3}\ln\left(\frac{(\mu+z)^2}{\mu^2+(\mu+z)^2}\right)\Big\}   
\end{eqnarray}
Similar to Eq.(\ref{bml}) and Eq.(\ref{bmg}) we have assumed $0\le z\le\mu$. Using Eq.(\ref{reltimesb}) but with the appropriate relaxation times determined in this section, we compute the resistance.
For this case $\Im[T_{k,k}(\epsilon^+)]=-\pi\sum_{k'}|T_{k,k'}(\epsilon^+)|^2\delta(\epsilon-\epsilon_{k'})$ and $|T_{k,k'}(\epsilon^+)|^2$ is independent of 
angle $\theta_{k,k'}$ between momenta $k$ and $k'$ since $|T_{k,k'}(\epsilon^+)|^2\propto|k|^2|k'|^2$.
The scattering rate to third order is 
\begin{eqnarray}\nonumber
\hbar\tau_{SP}^{-1}(\omega)&=&4 n_I S(S+1)\Big[\Big(\frac{2}{9 t^4}\Big)\Big(\frac{J}{2}\Big)^2|\omega+\mu|^3\\
&-&2\Big(\frac{2}{9 t^4}\Big)\Big(\frac{J}{2}\Big)^3|\omega+\mu|^3\Re[R_{SP}(\omega)]\Big]
\end{eqnarray}
The expression for the relaxation time $\tau_{SP}$,  within the same approach as the previous section, is
\begin{eqnarray}\label{relsp}
\hbar\tau_{SP}^{-1}(\omega)=4 n_I S(S+1)\Big(\frac{2}{9 t^4}\Big)\Big(\frac{J}{2}\Big)^2\frac{|\omega+\mu|^3}{1+J\Re[R_{SP}(\omega)]}
\end{eqnarray}

\begin{figure}[t]
\includegraphics[width=1\columnwidth, clip]{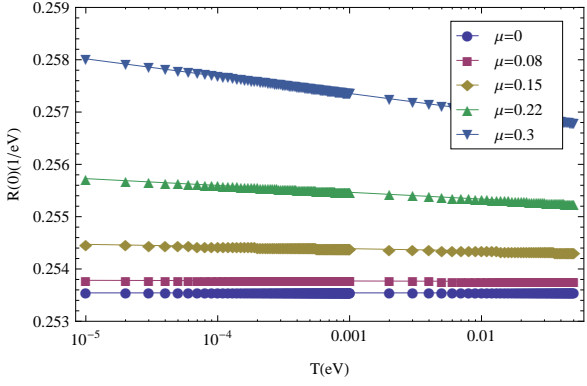}
\caption{The function $R_{SP}(0)$ plotted as a function of temperature $k_B T$ in the unit of $eV$. Energy cutoff $\Lambda=3.5eV$ and $t=2.7eV$.
Compared with Fig.\ref{fig2} $R_{SP}(0)$ shows more temperature variations when $T\rightarrow 0$, indicating the Kondo effect can only be observed on lower
temperature compared with symmetry breaking case.}
	  \label{fig3}
\end{figure}
The DC conductivity is related to the relaxation time with $\omega\rightarrow 0$. Fig.\ref{fig3} shows the function $R_{SP}(0)$ plotted as a function of temperature for different chemical potential. $R_{SP}(0)$ shows small variations with temperature except when temperature is close to zero where exponential growth with decreasing temperature is observed. For the case $\beta\mu\ll 2\pi$ the singularities from $1+J\Re[R_{SP}(0)]\simeq 0$ can be expressed , by using Eq.(\ref{bmml}), as
\begin{eqnarray}\nonumber
&&\frac{4J}{9\pi t^4}\left(\frac{\pi\Lambda^3}{\sqrt{2}}-\frac{1}{2\pi}\left(\frac{\pi}{\beta_k}\right)^3\right)\simeq -1\\\label{kltsp}
&&\rightarrow 
T_k\simeq \left(\frac{\sqrt{2}\Lambda^3}{\pi}\right)^{\frac{1}{3}}\left(1-\frac{J_c}{J}\right)^{\frac{1}{3}}
\end{eqnarray}
In above we have used the leading order correction as $(1/\beta)^3$ since its prefactor is $\mu/z$ which diverges as we take $z\rightarrow 0$. Higher order expansion in $\mu/z$ shows it as a sum of an infinite series in power of $(-\mu/z)^n$ with $n$ being some integer. Thus the infinite sum gives a factor of $-1$. 
\begin{figure}[t]
\includegraphics[width=1\columnwidth, clip]{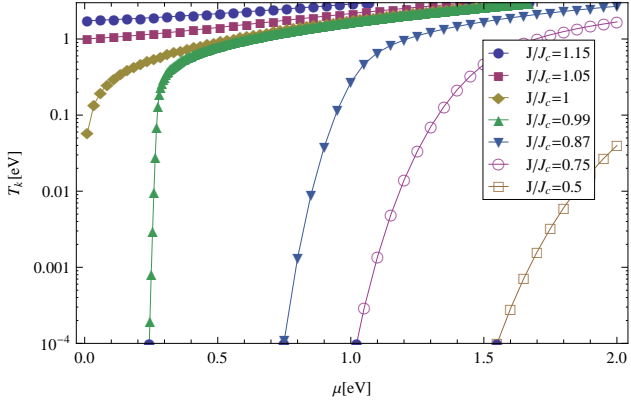}
\caption{The Kondo temperature as a function of chemical potential for various values of $J$. Energy cutoff $\Lambda=3.5eV$ and $t=2.7eV$ gives $|J_c|=\frac{9\sqrt{2}t^4}{4\Lambda^3}\simeq 3.94 eV$ for the symmetry preserving case.}
	  \label{fig4}
\end{figure}

In the opposite limit where $\beta\mu\gg 2\pi$ we use Eq.(\ref{bmmg})
\begin{eqnarray}\nonumber
&&\frac{4J}{9\pi t^4}\left(\frac{\pi\Lambda^3}{\sqrt{2}}+\frac{4}{\pi}\Lambda\mu^2+(1-\frac{8}{3\pi})\mu^3
\ln\left(\frac{\beta_k\mu}{2\pi}\right)\right)\simeq -1\\\label{khtsp}
&&\rightarrow  T_k\simeq \frac{\mu}{2\pi}\exp\left(c_2\left(\frac{\Lambda}{\mu}\right)^3\left(1-\frac{J_c}{J}\right)+c_3\frac{\Lambda}{\mu}\right)
\end{eqnarray}
where $c_2=\frac{3\pi^2}{\sqrt{2}(3\pi-8)}\simeq 14.69$, and $c_3=\frac{\pi^2}{4\sqrt{2}}\simeq 1.74$. Thus for both cases we obtain results similar to mean field results obtained in Ref.\onlinecite{Bruno}. 

Fig.\ref{fig4} shows the Kondo temperature as a function of chemical potential for various exchange coupling strength $J$. For $J/J_c$ smaller than $0.75$ the Kondo temperature is always smaller than the chemical potential for the range shown. $J/J_c$ approaches one from below and for $T_k\simeq \mu/2\pi$ the exponential dependence on $\mu$ crosses over to a power law. 

Given the relaxation time we obtain the linear response conductivity as
\begin{eqnarray}\label{cond}
\sigma_{SP}^s(T)&=&-\frac{2e^2}{3}\sum_n\int v_F^2\tau_{SP}(\epsilon_{k,n})\frac{\partial f}{\partial \epsilon_{k,n}}\frac{d\vec{k}}{(2\pi)^2}\\\nonumber
&=&-\frac{2e^2}{3}\frac{4v_F^2}{9\pi a_0^2t^2}\int_{-\infty}^{\infty}d\epsilon|\epsilon|\tau_{SP}(\epsilon-\mu)\frac{\partial f}{\partial \epsilon}
\end{eqnarray}
For $\beta\mu\ll 2\pi$, $T>\mu$, and $T>T_k$ we use Eq.(\ref{bmml}) and Eq.(\ref{kltsp}) and again approximate $\int g(\epsilon)(-\partial f/\partial \epsilon)\simeq \int_{\mu-T}^{\mu+T} g(\epsilon)/(2T)$. The conductivity for $0< \mu <T$ has no analytic form as the integral in Eq.(\ref{cond}) involves$\int_{\mu-T}^{\mu+T} d\epsilon/|\epsilon|^2$, which diverges as $\mu<T$. This vanishing resistivity for $\mu<T$ for impurities preserving the lattice symmetry is due to the fact that the scattering rate goes to zero faster than the chemical potential at the node. Thus the contribution to scattering near the node is dominated by other sources of scattering as compared to exchange scattering of impurities that preserve the lattice symmetry.

For $\beta\mu\ll 2\pi$, $\mu/2\pi<T<\mu$, and $T>T_k$ we use Eq.(\ref{bmml}) and Eq.(\ref{kltsp}) and use the same approximation scheme as above we obtain 
the resistivity as 
\begin{eqnarray}
\rho_{SP}^s(T)\simeq \frac{3}{2\pi}\frac{h}{e^2}n_I S(S+1)J\left(\frac{\mu^2-T^2}{T_k^3-T^3}\right)
\end{eqnarray}
Thus even for larger chemical potential the resistivity does not scale solely with Kondo temperature. The behavior for $T<\mu$ goes like $\mu^2/(T^3-T_k^3)$.

For large chemical potential or $\beta\mu\gg 1$ we approximate $-\partial f/\partial \epsilon\simeq \delta(\epsilon-\mu)$. Combining with Eq.(\ref{bmmg}) and Eq.(\ref{khtsp}) we get
\begin{eqnarray}
\rho_{SP}^s(T)\simeq \frac{9\pi^2}{(8\pi-8)}n_IS(S+1)\frac{h}{e^2}\frac{J}{\mu}\left(\ln\left[\frac{T_k}{T}\right]\right)^{-1}
\end{eqnarray}
Thus at large chemical potential the Kondo contribution to resistivity is similar in the form as the magnetic impurities in conventional metal.

\section{Resistivity due to impurity charge fluctuations}
From $|J|\simeq\frac{V^2|U|}{|(\epsilon_d-\mu)(\epsilon_d+U-\mu)|}\simeq V^2/|\mu-\epsilon_d|$, for large Coulomb repulsion $U$, it follows that to obtain  $|J|\geq |J_c|$ the impurity level $\epsilon_d$ must be close to the Fermi surface $\mu$. Since the density of state in the graphene is proportional to the energy
scale away from this Fermi surface, or $\rho_g(\epsilon)\propto |\epsilon|$ with $\rho_g(\epsilon)$ denoting graphene density of state, the phase space for charge fluctuation is very small and the local moment region is large compared with the case of the magnetic impurities in the conventional metal\cite{Bruno}. However it is still likely to have impurity level close to $\mu$ which is not in the local moment region\cite{Bruno2}. Thus it is worthwhile to estimate the resistivity contribution from impurity charge fluctuations.

 We use mean field approach on the Anderson impurity model shown in Eq.(\ref{Anderson}) and rewrite $H_U\rightarrow \sum_{s}Un_{-s}d_s^{\dagger}d_s-Un_{\uparrow}n_\downarrow$ with $n_s=\langle d_s^\dagger d_s\rangle$ determined self consistently, to obtain the impurity Green's function. From the imaginary part of this Green's function we obtain temperature dependence of the linear response resistivity by assuming Boltzmann transport. Under this mean field approach we obtain the retarded impurity Green's function  
as\cite{Bruno2}
\begin{eqnarray}\label{dotgr}
G^R_{dd,s}(\omega)=\frac{1}{\omega-\epsilon_d-Un_{-s}-\Sigma^R_{dd,s}(\omega)+i0^+}
\end{eqnarray}
The self energy part $\Sigma^R_{dd,s}(\omega)$ is given by 
\begin{eqnarray}\label{selfen}
\Sigma^R_{dd,s}(\omega)&=&\frac{V^2}{N}\sum_{\vec{q},n}|\Theta_{\vec{q},n}|^2G_{cc,s}^{0R}(\vec{q},\omega)\\\nonumber
&=&\frac{V^2}{N}\sum_{\vec{q}} \frac{|\Theta_q|^2(\omega+\mu)}{(\omega+\mu)^2-v_F^2|q|^2+i0^+sign(\omega+\mu)}
\end{eqnarray}
In above we have used $|\Theta_{\vec{q},n}|=|\Theta_{q}|^2$. $|\Theta_{q}|^2=1/2$ for symmetry breaking case and $|\Theta_{q}|^2=9|q|^2a_0^2/8$ for symmetry preserving case. We take the principal part of $\Sigma^R_{dd,s}(\omega)$ between $(-\Lambda,\Lambda)$ with $\Lambda$ being the linear spectrum cutoff. In the non-magnetic mixed valence regime, of which we are interested in, $0<n_s=n_{-s}< 1/2$. The impurity occupation $n_s$ is given by
\begin{eqnarray}\label{dotocp}
n_s=\int_{-\Lambda}^{\mu}d\omega \frac{-1}{\pi}\Im[G_{dd,s}^R(\omega)]
\end{eqnarray}
By using Eq.(\ref{dotgr}) and Eq.(\ref{selfen}) we find the relation between $\epsilon_d$ and $n_s$ by solving self-consistent conditions numerically.

\subsection{$C_{3v}$ symmetry breaking impurities}
For impurities breaking the symmetry the self energy $\Sigma^R_{dd,s}(\omega)$ obtained from Eq.(\ref{selfen}) is given by
\begin{eqnarray*}
\Sigma^R_{dd,s}(\omega)&=&-\frac{2V^2}{9\pi t^2}\Big[(\omega+\mu)\ln\left(\frac{|(\omega+\mu)^2-\Lambda^2|}{(\omega+\mu)^2}\right)\\&+&i\pi|\omega+\mu|\Big]
\end{eqnarray*}
Since $T_{k,k}(\omega)=\sum_s V^2|\Theta_k|^2 G_{dd,s}^R(\omega)$ we use Eq.(\ref{reltimesb}) and Eq.(\ref{sbcond}) to obtain the impurity conductivity, denoted as $\sigma^c_{SB}(T)$. The resistivity $\rho^c_{SB}(T)=1/\sigma^c_{SB}(T)$. We are mainly interested in the leading order temperature dependence of the resistivity contributed by the charge fluctuation in the Anderson impurity model. Thus we use the same approximation $-\int g(\epsilon)\partial f/\partial \epsilon\simeq \int_{\mu-T}^{\mu+T} g(\epsilon)/2T$ in Eq.(\ref{sbcond}) to extract the leading order in temperature dependence. The resistivity obtained for $0\simeq\mu<T$ is
\begin{eqnarray}
&&\rho^c_{SB}(T)\simeq\frac{9 n_I V^4}{t^2}\frac{h}{e^2}\frac{1}{r(\mu,T)}\\\nonumber
&&r(\mu,T)= 3 T^2 \alpha  \ln \left(\frac{\Lambda ^2-\tilde{\epsilon}_d^2}{T^2}\right)
   \Big(3 \alpha  \ln \left(\frac{\Lambda ^2-\tilde{\epsilon}_d
   ^2}{T^2}\right)+4 \alpha
\\\nonumber&& -6\Big)+T^2 \left(\alpha  \left(\left(8+9
   \pi ^2\right) \alpha -12\right)+9\right)+27 (\tilde{\epsilon}_d +\mu )^2   
\end{eqnarray}
Here $\alpha\equiv \frac{-2V^2}{9\pi t^2}$ and $\tilde{\epsilon}_d\equiv \epsilon_d+Un_{-s}$. Analytic result of resistivity for $T\le\mu$ can also be obtained but the expression are cumbersome and we defer a numerical analysis to section V. From Eq.(\ref{dotocp}) we find the non-magnetic region\cite{Bruno2} by demanding $n_s=n_{-s}$ when $\epsilon_d\simeq\mu$. Within this charge fluctuation regime ($0\le n_s\ll 0.5$) we study the temperature variation of resistivity at a given $\mu$ and Coulomb repulsion $U$.

\subsection{$C_{3v}$ symmetry preserving impurities}
For impurities preserving the honeycomb lattice symmetry the self energy $\Sigma^R_{dd,s}(\omega)$ is given by
\begin{eqnarray}\nonumber
\Sigma^R_{dd,s}(\omega)&=&-\frac{2V^2}{9\pi t^4}\Big[(\omega+\mu)^3 P\int_0^{\frac{\Lambda^2}{(\omega+\mu)^2}}\frac{x dx}{x-1}\\\label{sesp}&+&i\pi|\omega+\mu|^3\Big]
\end{eqnarray}
 We again use Eq.(\ref{reltimesb}) and Eq.(\ref{sbcond}) to obtain the impurity conductivity, denoted as $\sigma^c_{SP}(T)$. The resistivity $\rho^c_{SP}(T)=1/\sigma^c_{SP}(T)$. For temperature dependence we use the approximation $-\int g(\epsilon)\partial f/\partial \epsilon\simeq \int_{\mu-T}^{\mu+T} g(\epsilon)/2T$ in Eq.(\ref{sbcond}) to extract the leading order. To perform this computation we need to find the $\omega$ dependence  
in the principal integral of Eq.(\ref{sesp}). This is done by fitting numerically the principal value of the integral for large $\Lambda/(\omega+\mu)$. This
is because the relevant integration region for $\omega$ in the expression of conductivity is $\omega\subset(-T,T)$ which makes $\Lambda\gg |\omega+\mu|$ in our discussion. From the numerical fit with $10<\frac{\Lambda^2}{(\omega+\mu)^2}<10^2$ (chosen for experimentally accessible range) we have
\begin{eqnarray*}
P\int_0^{\frac{\Lambda^2}{(\omega+\mu)^2}}\frac{x dx}{x-1}\simeq 2.589+1.022\frac{\Lambda^2}{(\omega+\mu)^2}
\end{eqnarray*}
The conductivity obtained is
\begin{eqnarray*}
&&\sigma^c_{SP}(T)\simeq\frac{2e^2}{3}\frac{4v_F^2}{9\pi a_0^2t^2}\int_{\mu-T}^{\mu+T}d\epsilon|\epsilon|\frac{\tau^c_{SP}(\epsilon-\mu)}{2T}\\
&&\frac{1}{\tau^c_{SP}(\epsilon-\mu)}=-\frac{2n_I}{\hbar}\frac{V^2|\epsilon|^2}{2t^2}\sum_s\Im[G_{dd,s}^R(\epsilon-\mu)]
\end{eqnarray*}
For $0<\mu\le T$ the resistivity $\rho^c_{SP}(T)=1/\sigma^c_{SP}(T)\rightarrow 0$ similar to the case for spin fluctuation in Eq.(\ref{cond}).
For $T<\mu$ we have
\begin{eqnarray}
\rho^c_{SP}(T)&=&\frac{27\pi\alpha}{8}\frac{h}{e^2}n_I\frac{3\pi T\alpha(T^2-\mu^2)^5}{t^2r^c_{SP}(\mu,T)}\\\nonumber
r^c_{SP}(\mu,T)&=&T\left((a^4+\pi^2)\alpha^2T^{12}/t^4+\ldots\right)\\\nonumber&+&3a^2\frac{\alpha}{t^2}(\tilde{\epsilon}_d+\mu)(\mu^2-T^2)^5\ln(\frac{\mu-T}{\mu+T})
\end{eqnarray}
Here $a=2.589$ is the parameter from principal integral of the dot self energy. We find the non-magnetic region from Eq.(\ref{dotocp}) and study the temperature variation of resistivity within this charge fluctuation regime ($0\le n_s\ll 0.5$).
 
\subsection{Range of validity for mean field result} 
\begin{figure}[t]
\includegraphics[width=1\columnwidth, clip]{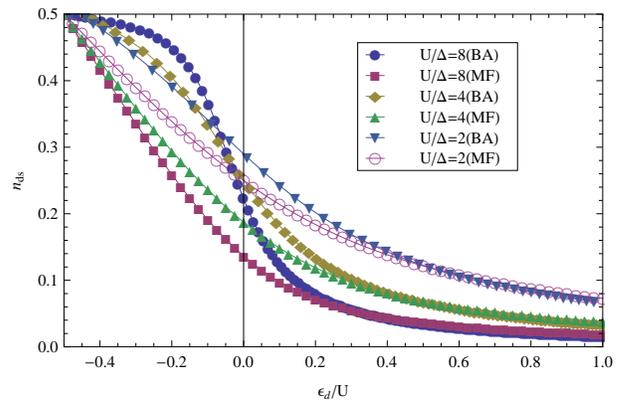}
\caption{Comparison of Bethe Ansatz with mean field d level occupation vs impurity level for conventional metal. Fermi energy is set at $\epsilon_d=0$. Lines denoted BA means Bethe Ansatz results and lines denoted MF means mean field results. The two match better in the region where $n_{ds}\le 0.1$. This upper bound increases with decreasing $U/\Delta$.}
	  \label{fig5}
\end{figure}

Before we proceed to compare the  temperature dependences of resistivity due to charge fluctuations, spin fluctuations, and the influence of impurities position, we pause here to discuss the regimes where mean field results are valid in this Anderson impurity model. The order parameter of this unrestricted Hatree Fock is the d level occupation for a given spin $n_{d,s}$. To make comparison with exact d level occupation done by Bethe Ansatz\cite{Tsvelik} we need to go back to the case for conventional metal where the mean field results were done by P. W. Anderson\cite{Anderson}.
The d level occupation for a given spin $s$ at zero temperature is 
\begin{eqnarray}\label{dot}
n_{d,s}=\frac{1}{\pi}\cot^{-1}\left(\frac{\epsilon_d+U n_{d,-s}-\epsilon_F}{\Delta}\right)
\end{eqnarray}
Here $\Delta=\pi V^2\rho(\epsilon_F)$ with $\rho(\epsilon_F)=1/2\pi$ the density of state for conventional metal. We solve Eq.(\ref{dot}) in the non-magnetic region where $n_{d,s}=n_{d,-s}$ and compare the answers with exact results obtained by Bethe Ansatz. The comparison for d level occupation for a given spin v.s. impurity level, with $\epsilon_F=0$, is shown in Fig.\ref{fig5}.   
From this figure we can see that the mean field results deviate from exact ones in a range $n_{d,s}\sim 0.06-0.1$ for the range of $U\sim 2-8\Delta $, indicating that mean field is a good approximation when the impurity level $\epsilon_d$ is higher than the Fermi energy $\epsilon_F$ or, in the other words, the impurity is nearby the empty orbital region. For larger Coulomb repulsion $U/\Delta$ the minimum value of $\epsilon_d$ of overlapping region is closer to $\epsilon_F$. Since for two dimensional system the mean field results are marginal, we expect the mean field result work for $\epsilon_d\simeq\epsilon_F=\mu$, as the s wave scattering in the conventional metal considered above\cite{Tsvelik,Anderson} is a one dimensional problem. Since the crossover shifts to lower and lower values of $n_{d}$ as $U$ increases, this is a rough criterion but establishes a basis for the mean field calculations.

\section{Resistivity temperature dependence}
In sections III and IV we have shown the analytic results of temperature dependence of resistivity for $\beta\mu\gg 1$ and $\beta\mu\ll 1$.
Here we compute numerically the temperature dependence of resistivity due to spin fluctuations, $\rho^s_{SB}(T)$ and $\rho^s_{SP}(T)$ for impurities breaking/preserving honeycomb lattice symmetry, and the temperature dependence of resistivity due to charge fluctuations, $\rho^c_{SB}(T)$ and $\rho^c_{SP}(T)$.
We use the full form of $R_{SB}(\omega)$ and $R_{SP}(\omega)$ and extract the results for $T>T_k$ with $T_k$ obtained numerically the same way as we obtain the Kondo temperature in Fig.\ref{fig2} and Fig.\ref{fig4}. We compare the resistivity for different symmetry with the same sets of parameters. The resistivity due to symmetry preserving impurities is much smaller than that of symmetry breaking case due to the factor of $(\mu/t)^2$ (see Eq.(\ref{relsb}) and Eq.(\ref{relsp})). We examine the resistivity due to impurities spin and charge fluctuations in the symmetry breaking case and make comparison with the experimental results\cite{Chen} in the next section.
\subsection{Comparison of resistivity due to spin and charge fluctuations with different symmetry} 
We use $t=2.7eV$, $\Lambda=3.5eV$, $V=1eV$, and $U=4eV$ in all of the numerical results within this section. We choose different impurity level $\epsilon_d$ to explore the resistivity due to spin and charge fluctuations. The resistivity v.s. temperature is evaluated numerically between $T\subset(10^{-4},10^{-1})eV$.

\begin{figure}[t]
\includegraphics[width=1\columnwidth, clip]{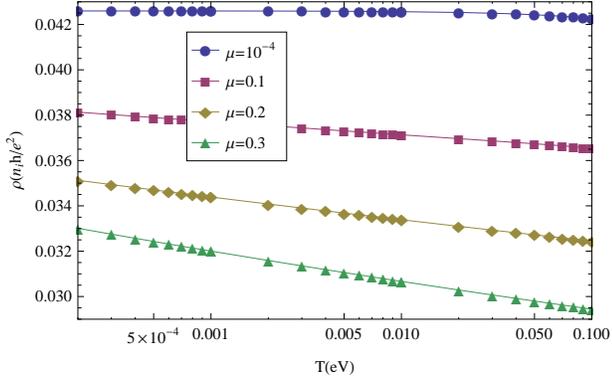}
\caption{Perturbation results for resistivity v.s. temperature for symmetry breaking case with $\epsilon_d=-1$ and various $\mu$. The exchange coupling strength
$|J|$ is larger than $0.2432$ (for $\mu=0.3eV$) and less than $0.2884eV$ (for $\mu=10^{-4} eV$) in the range chosen. In all cases the Kondo temperature is less than $10^{-12}eV$ so the temperature range chosen is well above the Kondo scale.}
	  \label{fig6}
\end{figure}

\begin{figure}[t]
\includegraphics[width=1\columnwidth, clip]{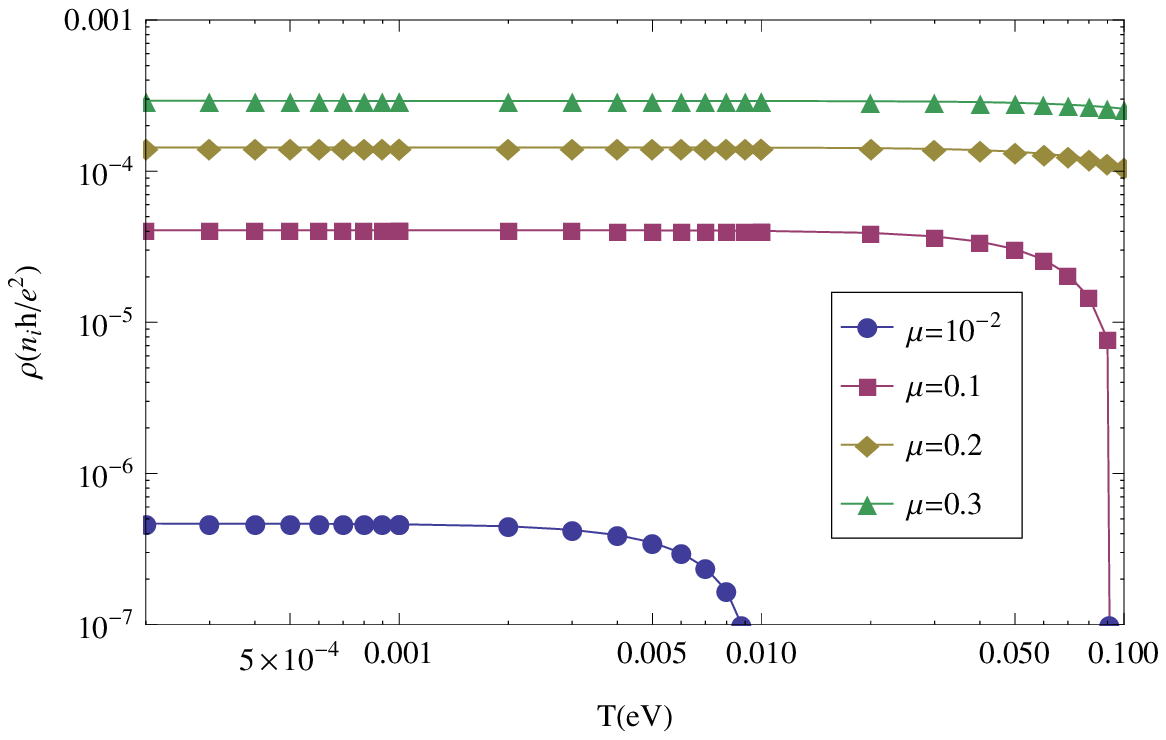}
\caption{Perturbation results for resistivity v.s. temperature for symmetry preserving case with $\epsilon_d=-1$ and various $\mu$. The exchange coupling strength
$|J|$ is larger than $0.2889$ (for $\mu=0.3eV$) and less than $0.3380eV$ (for $\mu=10^{-2} eV$) in the range chosen. In all cases the Kondo temperature is less than $10^{-12}eV$ so the temperature range chosen is well above the Kondo scale.}
	  \label{fig7}
\end{figure}

Let us first study the local moment region. We choose $\epsilon_d=-1eV$ to ensure the d level occupation $n_{d,s}\simeq 0.5$. The chemical potential $\mu$ is chosen between $10^{-4}eV$ to $0.3eV$. From $|J|\simeq\frac{V^2|U|}{|(\epsilon_d-\mu)(\epsilon_d+U-\mu)|}$ this choice of parameters renders the exchange coupling strength $1.14eV<|J|<1.33eV$. For both cases these exchange coupling strengths are less than the critical value $|J_c|$ and the Kondo temperature $T_k$ obtained for both cases are extremely small ($T_k<10^{-12}eV$). With this choice of parameters $T_k\ll\mu$ and therefore the analytic expression for $T_k$ corresponds to Eq.(\ref{logtk}) for symmetry breaking case and Eq.(\ref{khtsp}) for symmetry preserving case. The resistivity v.s. temperature are plotted in Fig.\ref{fig6} and Fig.\ref{fig7}. 
\begin{figure}[t]
\includegraphics[width=1\columnwidth, clip]{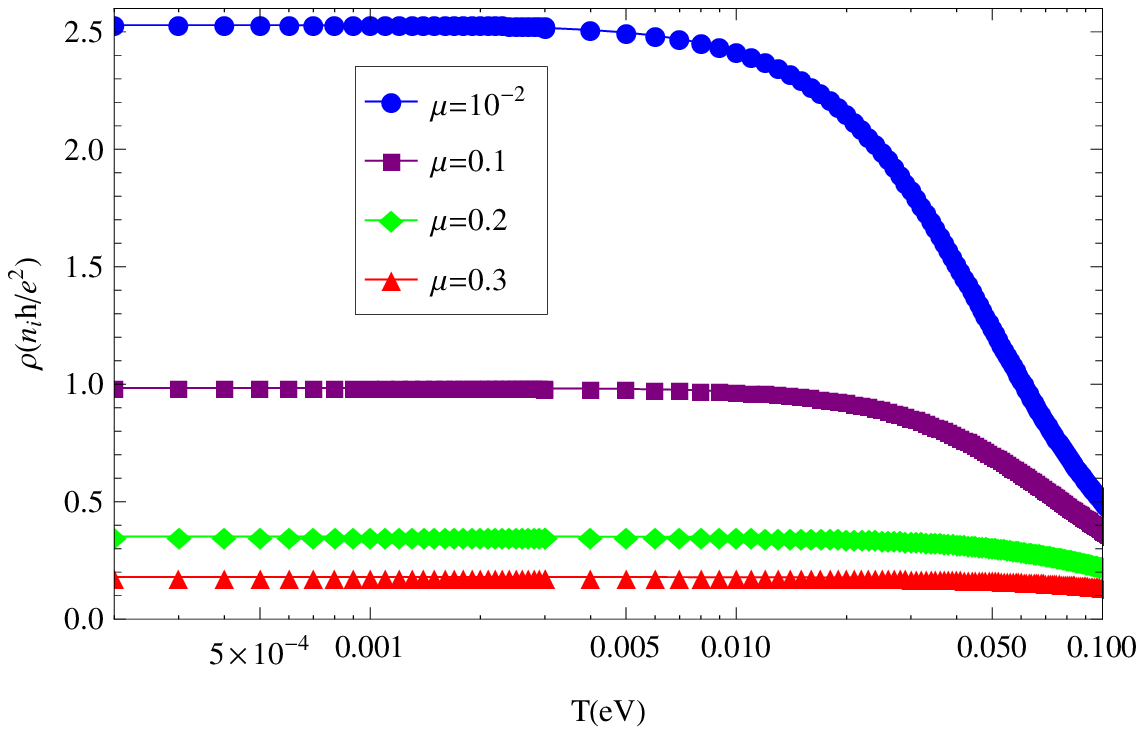}
\caption{Mean field results for resistivity v.s. temperature for symmetry breaking case with $\epsilon_d=0$ and various $\mu$. The d level occupation $n_{d,s}$ is greater than $0.024$ (when $\mu=10^{-2} eV$) and lesser than $0.093$ (when $\mu=0.3 eV$) in the chosen range.}
	  \label{fig8}
\end{figure}
\begin{figure}[t]
\includegraphics[width=0.9\columnwidth, clip]{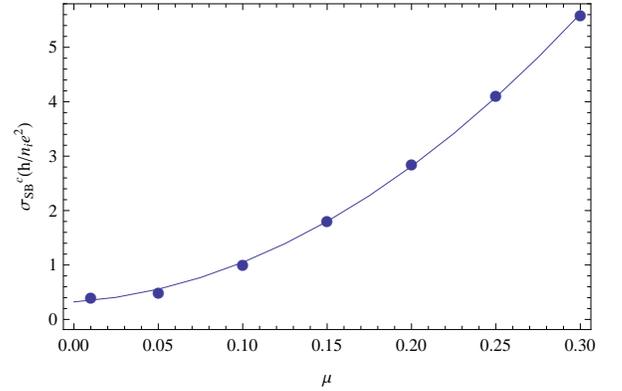}
\caption{Mean field results for conductivity at $T=10^{-4}eV$ v.s. chemical potential $\mu$ (in unit of $eV$). Dots are the numerical data of conductivity at various chemical potential and the line is the fitted quadratic curve with $\sigma\simeq 0.32+2.08\mu+51.86\mu^2$. This quadratic behavior is also seen in Fig.3d of Ref.\onlinecite{Chen}.}
	  \label{fig8-1}
\end{figure}
\begin{figure}[t]
\includegraphics[width=1\columnwidth, clip]{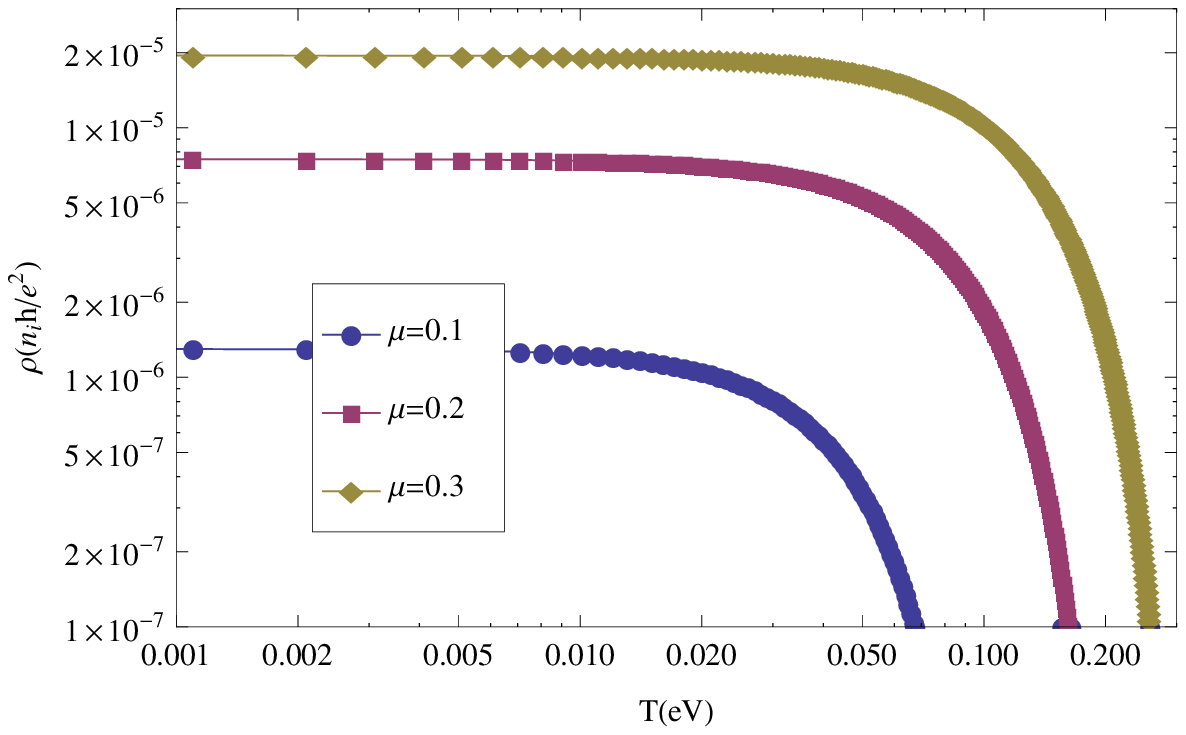}
\caption{Mean field results for resistivity v.s. temperature for symmetry preserving case with $\epsilon_d=0$ and various $\mu$. The d level occupation $n_{d,s}$ is greater than $0.040$ (when $\mu=0.1 eV$) and lesser than $0.093$ (when $\mu=0.3 eV$) in the chosen range.}
	  \label{fig9}
\end{figure}

In Fig.\ref{fig6} we see the $T_k\ll T$ tails of the logarithmic upturns occurring when $T\simeq T_k$. The resistivity goes down as chemical potential increases.
This tendency is quite different from the case of symmetry preserving ones, shown in Fig.\ref{fig7}. The dependence of resistivity on chemical potential $\mu$ for symmetry preserving case shows $\rho_{SP}^s(\mu)\propto\mu^2$ by comparing the resistivity at $T=10^{-4} eV$ in Fig.\ref{fig7}. At temperature higher than the chemical potential the resistivity goes down with increasing temperature faster than the logarithmic tail for all cases in Fig.\ref{fig7}. This is due to the divergence in conductivity when $\mu\simeq T$. The order of magnitude of resistivity at the same temperature for symmetry preserving case is much smaller than the resistivity for symmetry breaking case. Thus we can safely ignore the contributions from symmetry preserving type of impurities when considering the resistivity due to spin fluctuations.

 For the case of charge fluctuations we choose $\epsilon_d=0eV$ to ensure the d level occupation $n_{d,s}< 0.1$. The chemical potential $\mu$ is chosen between $10^{-4}eV$ to $0.3eV$. We compute the resistivity v.s. temperature for $T\subset (10^{-4},10^{-1})eV$ numerically from the mean field results. The resistivity v.s. temperature are plotted in Fig.\ref{fig8} and Fig.\ref{fig9} for symmetry breaking and symmetry preserving cases.

Fig.\ref{fig8} shows $\rho_{SB}^c(T)\propto\ln(T)$ for $T>10^{-2}eV$ and tends to a flat region for small temperature, which is very similar to the screening result of Kondo effect at $T<T_k$. Conductivity at $T=10^{-4}$ shows quadratic chemical potential dependence, shown in Fig.\ref{fig8-1}, consistent with the gate voltage dependence on conductivity seen in the experiment\cite{Chen}. The experimental fit in Ref.\onlinecite{Chen} for Kondo scale, however, is about one order of magnitude smaller compared with  the energy scale obtained in logarithmic temperature range in Fig.\ref{fig8}.  The temperature dependence of resistivity in this charge fluctuation regime is similar to that of Kondo model in this case but the physics is not related to spin but charge fluctuation. To facilitate comparing our results with experimental ones in Ref.\onlinecite{Chen} we refer to the energy scale as a Kondo-like temperature in the following discussion.

 Fig.\ref{fig9} also shows $\rho_{SP}^c(T)\propto\ln(T)$ for $T\le\mu$ with shorter range of temperature and similarly tends to a flat region for small temperature. The resistivity increases with increasing chemical potential in Fig.\ref{fig9} similar to the case of spin fluctuation. The resistivity for symmetry preserving case is much smaller than that for symmetry breaking impurities and thus ignore the contribution from symmetry preserving impurities.

In summary when both types of impurities are present the resistivity due to impurities preserving lattice symmetry is much smaller than that from impurities breaking the symmetry. Thus we focus our discussions on symmetry breaking cases for spin and charge fluctuations in the next section.

\section{Comparison with experimental data}
Here we make comparisons with the experimental data given in Ref.\onlinecite{Chen}. We start with the perturbative results of Kondo model in the case of impurities breaking the honeycomb symmetry. To have large Kondo temperature ($30K\le T_k\le 90K$ in the experiment) the exchange coupling $|J|$ must be very close to its critical value $|J_c|$. As perturbation breaks down when $T\sim T_k$, we can only analyze the gate voltage dependence of Kondo temperature shown in Fig.4 of Ref.\onlinecite{Chen}. The strategy is the following: We find the impurity level $\epsilon_d$ at a given chemical potential $\mu$ by using the experimental Kondo temperature $T_k$ as the Kondo temperature obtained by the pole of resistivity, or $1+J R_{SB}(0)=0$ where $J\simeq V^2/(\epsilon_d-\mu)(\epsilon_d+U-\mu)$. 

In the experiment the Kondo temperature is obtained as a function of gate voltage. We assume the gate voltage $V_g$ is connected with chemical potential $\mu$ via capacitive effect, i.e. $Q/e=c_g V_g/e=8c_{g}\mu^2/(27\sqrt{3}\pi t^2 a_0^2)$ with $Q$ representing the electric charges, $e=1.6\times 10^{-19}C$, and $c_g=1.15\times 10^{-8} F/cm^2$ as the capacitance of the graphene. In the experiment of Ref.\onlinecite{Chen} $V_g=5.3V$ is regarded as the position of the Dirac node. Thus we take $V_g=5.3+8e/(27\sqrt{3}\pi t^2a_0^2c_g)\mu^2=5.3+515.387\mu^2$ by fixing $V_g=5.3V$ at $\mu=0eV$. Using the experimental Kondo temperature at a given chemical potential we compute the corresponding exchange coupling strength $J$ and thus determine the relationship between $\mu$ and impurity level $\epsilon_d$. The results are shown in Table.\ref{table1}

\begin{table}
\begin{center}
\begin{tabular}{| c || c || c|| c | }
  \hline                       
  $T_k$(K) & $V_g$(V) & $\mu$(eV) & $\epsilon_d$(eV) \\ \hline
  31.5 & 5.3 & 0 & -0.225949 \\ \hline
  32 & 6 & 0.0368538 & -0.191582 \\ \hline
  35 & 10 & 0.0954953 & -0.140059 \\ \hline
  40 & 12.5 & 0.118195 & -0.119956\\ \hline
  51 & 15 & 0.137189 & -0.102273\\ \hline
  56.2 & 20 & 0.168885 & -0.0743727\\
  \hline   
\end{tabular}
\caption{Relation between $\mu$ and $\epsilon_d$ obtained by fitting with experimental Kondo temperature. The left two columns show experimental Kondo temperature at a given gate voltage. We compute corresponding chemical potential in the third column by using $V_g=5.3+515.387\mu^2$. The impurity level $\epsilon_d$, shown in the last column, is obtained by evaluating the corresponding exchange coupling stregth $J$.} \label{table1}
\end{center}
\end{table}

\begin{figure}[t]
\includegraphics[width=0.9\columnwidth, clip]{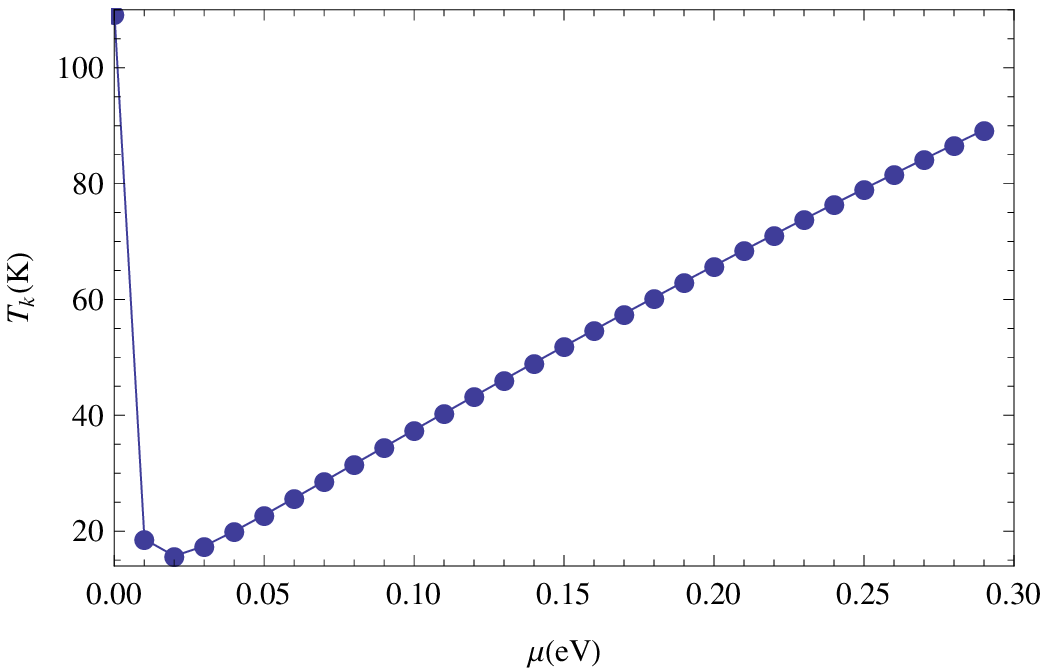}\\
\includegraphics[width=0.5\columnwidth, clip]{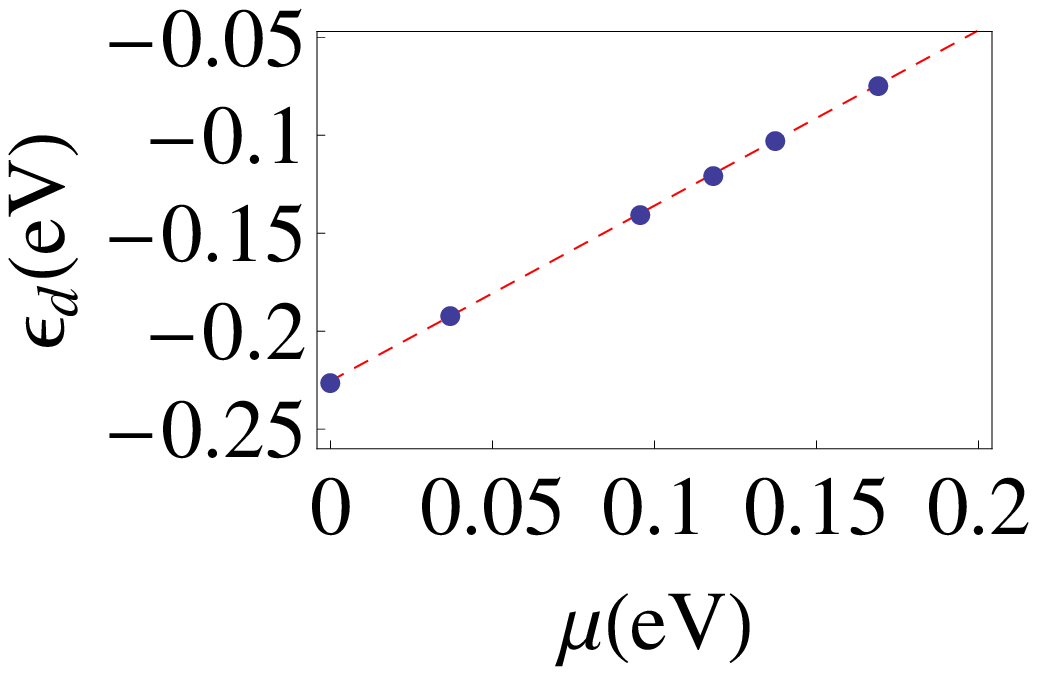}\hfill
\includegraphics[width=0.5\columnwidth, clip]{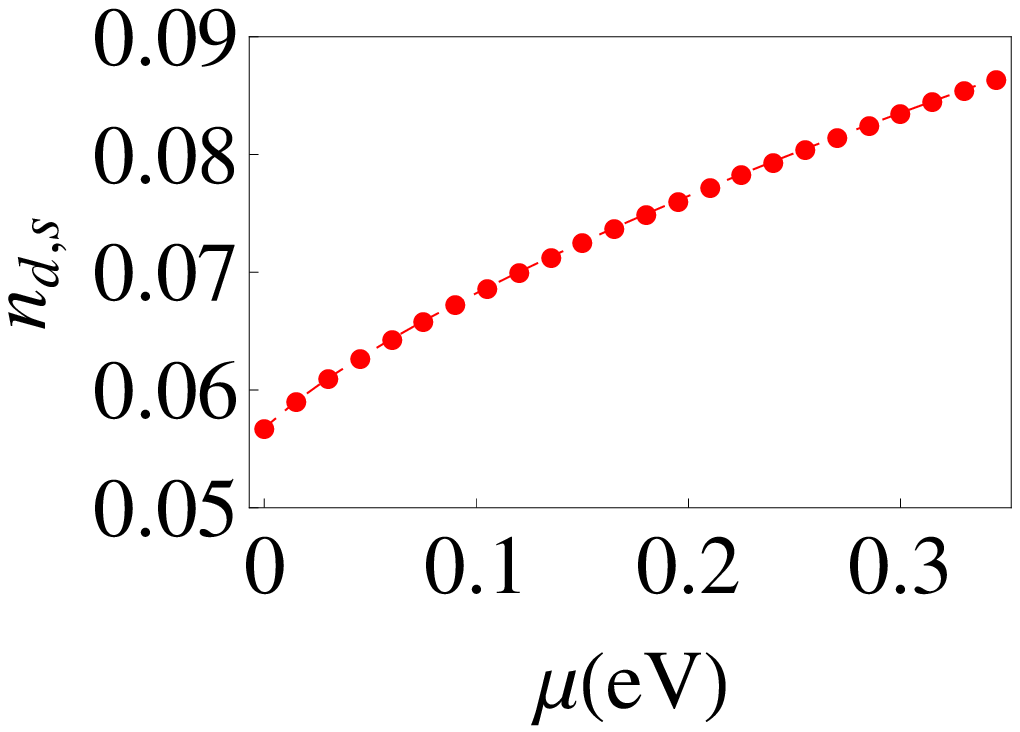}
\caption{Top:Kondo temperature $T_k$ as a function of chemical potential $\mu$. The relationship is obtained by $|J|\sim V^2/|\mu-\epsilon_d|$ and $\epsilon_d\propto\mu$ which we find by fitting experimental data shown in the lower left figure. Lower left: Impurity level $\epsilon_d$ as a function of chemical potential. Dots are obtained by the data in Fig.4 of Ref.\onlinecite{Chen} shown in Table.\ref{table1}. Red dashed line is the linear fitting function which gives $\epsilon_d=-0.2254+0.8951\mu$. Lower right: Impurity occupation of a given spin $n_{d,s}$ as a function of chemical potential evaluated by mean field approach.}
	  \label{fig11}
\end{figure}
\begin{figure}[t]
\includegraphics[width=0.9\columnwidth, clip]{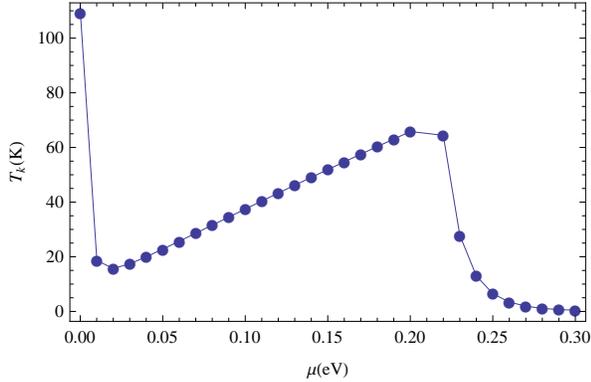}
\caption{Kondo temperature $T_k$ as a function of chemical potential $\mu$. The relationship is obtained by $|J|\sim V^2/|\mu-\epsilon_d|$ and $\epsilon_d=-0.2254+0.8951\mu$ for $0\le\mu\le 0.21eV$ and $\epsilon_d=-0.029456$ (corresponds to $V_g=30V$ or $\mu=0.2189eV$) for $0.21 eV<\mu\le 0.3 eV$}
	  \label{fig11p}
\end{figure}
From Table.\ref{table1} we find $\epsilon_d\propto \mu$.
The obtained impurity level $\epsilon_d$ changes linearly with the chemical potential $\mu$ as shown in lower left of Fig.\ref{fig11}. One of the main conclusions of this work is that the observed upturn in resistivity\cite{Chen} can be understood in terms of an Anderson impurity model only if the impurity level varies with the applied voltage. By using the linear fit in this figure we obtain the Kondo temperature as a function of chemical potential shown in the top of Fig.\ref{fig11}. Between $\mu=0.02eV$ to $0.3eV$ the Kondo temperature grows monotonically from $14K$ to $90K$. The decrease of $T_k$ with increasing $\mu$ for $0eV<\mu<0.02eV$ may indicate the failure of linearity between $\mu$ and $\epsilon_d$ for the onset of nonzero chemical potential or the failure of the Kondo physics near the node. The chemical potential dependence shown in top figure of Fig.\ref{fig11} is roughly consistent with Fig.4 in Ref.\onlinecite{Chen} in the intermediate gate voltage. 

For gate voltage $V_g$ larger than $30V$ in Ref.\onlinecite{Chen}, the experimental $T_k$ begins to decrease with increasing gate voltage. This can be accounted for qualitatively, as shown in Fig.\ref{fig11p}, by assuming that the energy of the  impurity level no longer changes with the external gate voltage for $V_g>30V$ due to sufficient charge screening. For small gate voltage (chemical potential close to the node) the experimental $T_k$ increases monotonically with increasing gate voltage. In this region neither constant impurity level nor $\epsilon_d\propto\mu$ gives the corresponding experimental dependence on $V_g$ based on our perturbative Kondo results.  

We also compute the impurity occupation as a function of $\mu$ by using mean field as shown by Eq.(\ref{dotocp}) in symmetry breaking case. The obtained impurity occupation for a given spin increase from $0.057$ to $0.082$ monotonically between $\mu\subset(0,0.3)eV$. Given that the validity of the mean field is limited to small values of the impurity level occupations (see section IVC ), we expect deviations away from the mean field. Thus the system is not likely to stay in the local moment region near the node, suggesting  a cross over of impurity occupation from local moment to empty orbital regime based on numerical renormalization group results in Ref.\onlinecite{Carlos}. Thus Kondo effect alone would not be able to explain the logarithmic temperature dependence seen in Ref.\onlinecite{Chen}.

\begin{figure}[t]
\includegraphics[width=1\columnwidth, clip]{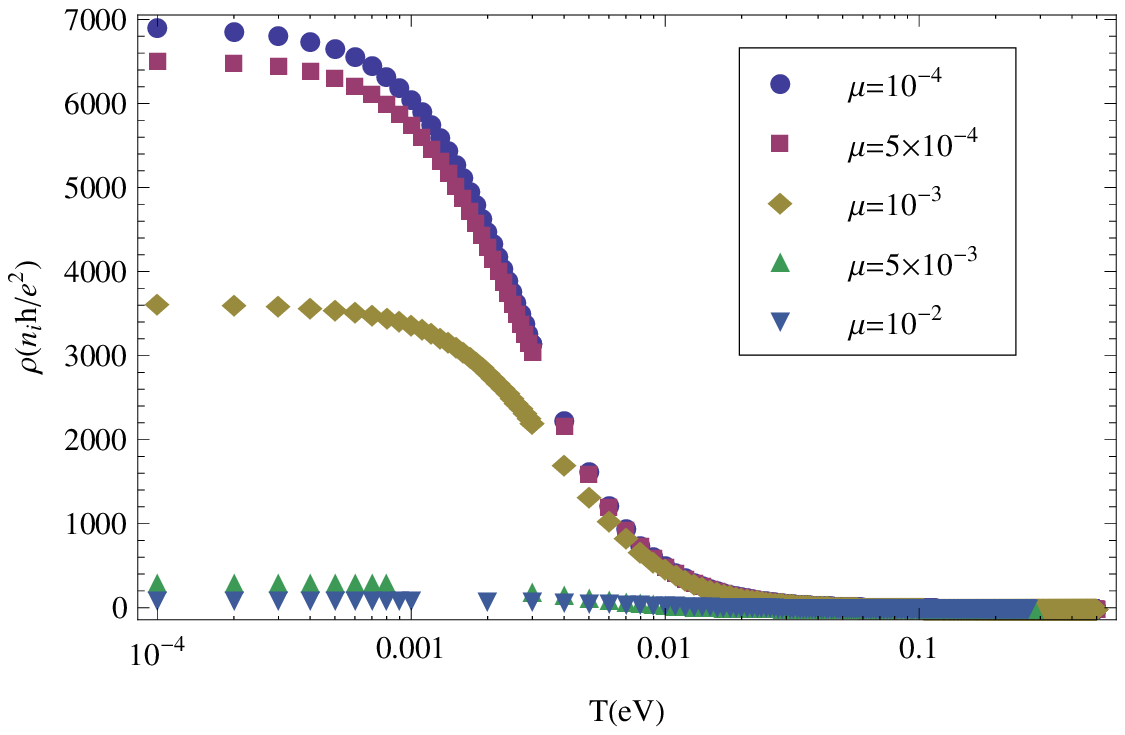}\\
\includegraphics[width=0.5\columnwidth, clip]{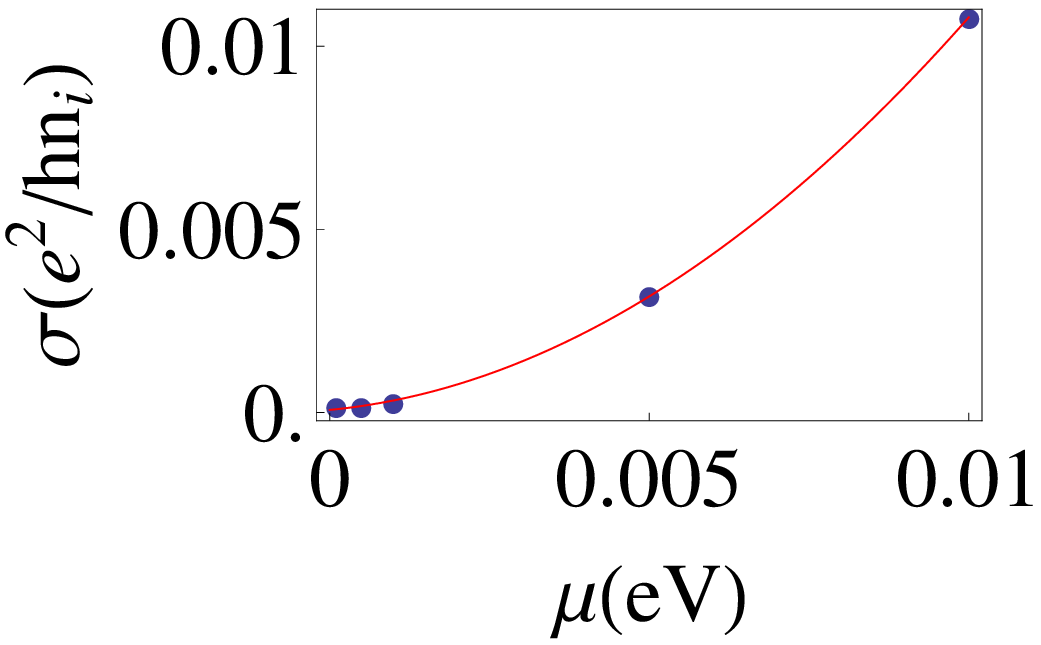}\hfill
\includegraphics[width=0.5\columnwidth, clip]{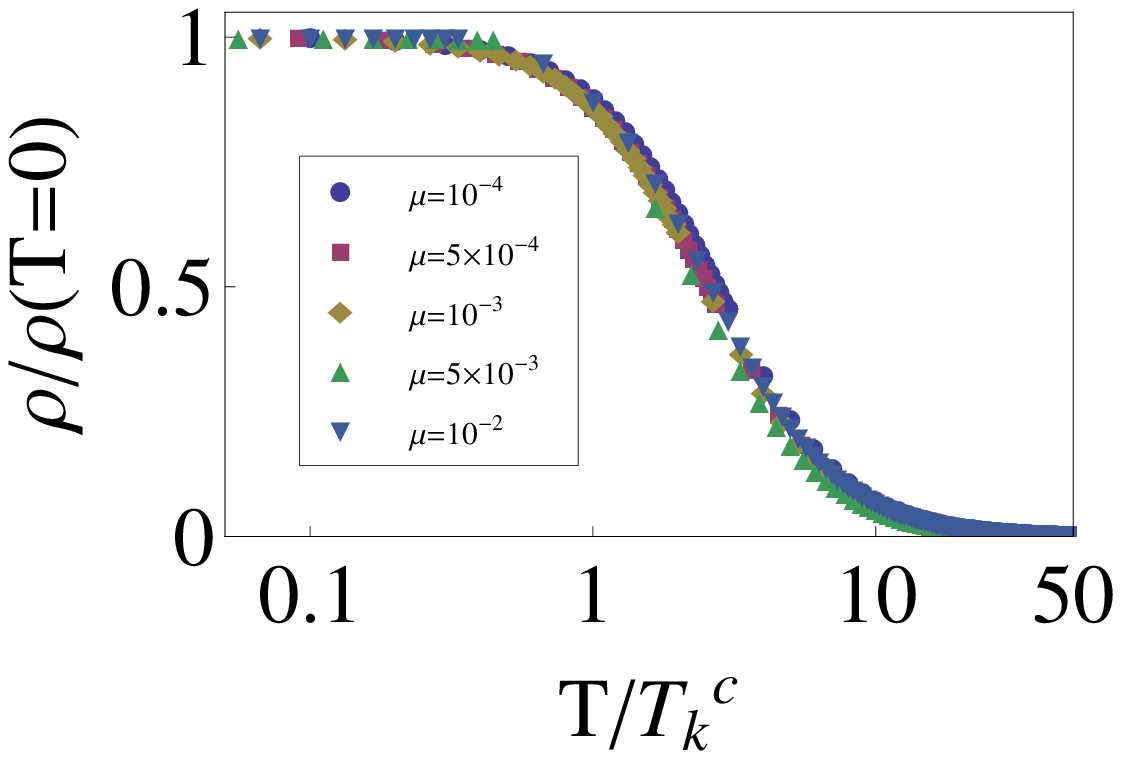}
\caption{Top: Resistivity v.s. temperature for charge fluctuation case with $\epsilon_d=-0.2254+0.8951\mu$ and chemical potential between $10^{-4}eV$ and $10^{-2} eV$. Lower left: Quadratic dependence on $\mu$ for conductivity at zero temperature. Red line is the quadratic fitting which gives $\sigma_{SB}^c(\mu,T=0)=0.0000693+0.1675\mu+90.3726\mu^2$ and blue dots are conductivity at various chemical potential. Lower right: Universality curve after rescaling resistivity and Kondo-like temperature scale $T_k^c$ obtained by the temperature scale when logarithmic behavior is shown. $1.1 T_k^c\simeq T_k$ by comparing this universality curve with Fig.2b in Ref.\onlinecite{Chen}.}
	  \label{fig13}
\end{figure}
\begin{figure}[t]
\includegraphics[width=1\columnwidth, clip]{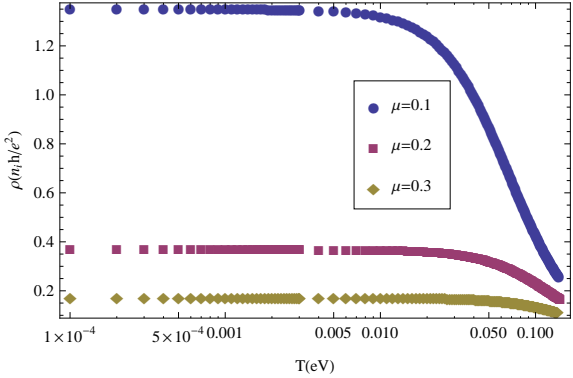}
\caption{Resistivity due to charge fluctuations v.s. temperature plot. For $T<10^{-2}eV$ the resistivity decreases slowly with temperature. For $10^{-2}eV<T$ the logarithmic dependence on temperature begins to appear.}
	  \label{fig12}
\end{figure}
Let us now investigate whether charge fluctuations can give rise to temperature dependence of resistivity seen in the experiment. We take  $\epsilon_d=-0.2254+0.8951\mu$ and evaluate the resistivity v.s. temperature from mean field results of impurity Green's function for symmetry breaking case.  For chemical potential close to the node, we get reasonable temperature scale (the logarithmic behavior shows up at $T\simeq 10^{-3}eV$) from charge fluctuations as shown in the top figure of Fig.\ref{fig13}. We also have $\mu^2$ being proportional to conductivity at zero temperature, as seen in the lower left of Fig.\ref{fig13} which was observed in the Ref.\onlinecite{Chen}. Rescaling $\rho(T)$ by $\rho(0)$ and $T$ by Kondo-like temperature $T_k^c$ obtained by the
temperature at which the resistivity begin to show logarithmic dependence in $T$, we obtain the universal curve shown in the lower right of Fig.\ref{fig13}. In this range of chemical potential the impurity occupation $n_{d,s}\simeq 0.057$. It shows that even for the chemical potential $\mu$ close to the node the one parameter scaling is still possible in this charge fluctuation scheme, while it is shown analytically in Eq.(\ref{sbr}) the one parameter scaling is unlikely for $\mu\simeq 0$
in the Kondo case.    

For chemical potential away from the Dirac node we plot the resistivity v.s. temperature for $\mu=0.1$, $0.2$ and $0.3eV$ in Fig.\ref{fig12}. The overall feature is very similar to the Kondo results: near zero temperature the resistivity decreases with $T^2$ while at large temperature $\rho(T)\propto\ln(T)$. At zero temperature the conductivity is proportional to $\mu^2$ as in the case shown in Fig.\ref{fig8-1}. However the logarithmic behavior shows up at $T\simeq 10^{-2} eV$ which is about one order of magnitude larger than the experimental results in Ref.\onlinecite{Chen}. Thus the charge fluctuation cannot explain the experiment for large chemical potential. 

\begin{figure}[t]
\includegraphics[width=1\columnwidth, clip]{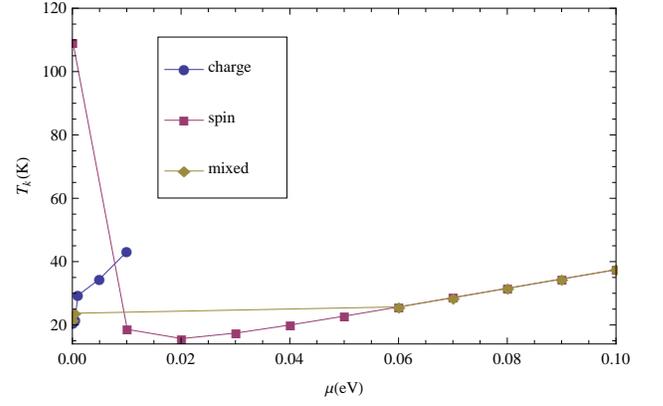}
\caption{Kondo temperature $T_k$ as a function of chemical potential $\mu$. Blue dots are the Kondo-like temperature, obtained by comparing universal curve with NRG results, for charge fluctuations case. Purple dots are the Kondo temperature for spin fluctuation case. Between $\mu=10^{-3}eV$ to $\mu=0.02eV$ both cases give Kondo temperature higher than their neighboring chemical potentials. The "mixed" (brown dots) case takes Kondo scale obtained by charge fluctuations with $\mu<10^{-3}eV$ and spin fluctuations with $\mu>0.02eV$. By combing these two the Kondo scale obtained grows monotonically with chemical potential.}
	  \label{fig14}
\end{figure}
 
By comparing the universal curve obtained by numerical renormalization group\cite{Costi} shown in Fig.2b in Ref.\onlinecite{Chen} and the one we have for charge fluctuations in lower right of Fig.\ref{fig13} we get $1.1 T_k^c\sim T_k$. The "Kondo temperature" for both charge and spin fluctuations as
a function of chemical potential is shown in Fig.\ref{fig14}. From Fig.\ref{fig14} we observe that charge fluctuations give large Kondo scale with increasing chemical potential and a good agreement with experimental results is obtained only if $\mu\le 10^{-3}eV$. Away from the node the Kondo scale obtained by charge fluctuations grows much faster than the that of the spin fluctuations. The Kondo scale obtained from spin fluctuation, on the other hand, gives large Kondo scale for $\mu<10^{-2}eV$ and reaches its minimum when $\mu\sim 0.02eV$. The combined picture of the two cases as shown in Fig.\ref{fig14}, by assuming charge fluctuation for $\mu\sim 0$ and spin fluctuation for large $\mu$, can give the overall consistent picture as seen in the experiment for gate voltage less than $30V$. For $10^{-3} eV<\mu<0.06 eV$ it shows the cross over from charge fluctuations to spin fluctuations, which is not accounted for in our simple mean field in Anderson model nor perturbation in Kondo model.  For gate voltage larger than $30V$ a non-monotonic dependence of $T_k$ on $\mu$ seen in Ref.\onlinecite{Chen}. Given our analysis we speculate that the screening due to the finite density of carriers could modify the dependence of the energy of the impurity level on the gate voltage. A weaker dependence at large gate voltage will lead to a decreasing Kondo temperature.

\section{Conclusion}
We use Anderson impurity model to describe the dilute impurities behavior in graphene. The goal is to test whether the recent experiment on the resistivity of graphene with vacancies induced by ion irradiation in ultra-high vacuum\cite{Chen} can be solely explained by the single impurity Kondo effect (spin fluctuations). To study this local moment regime we use Schrieffer Wolf transformation to freeze the charge degree of freedom and obtain the Kondo Hamiltonian. In the case of dilute impurities we may ignore the RKKY interactions and treat the problem as single impurity Kondo model.   

We have computed this Kondo contribution to DC resistivity by perturbation in $T$-matrix formulation. Analytic expressions are obtained for $\beta\mu\ll 2\pi$ and $\beta\mu\gg 2\pi$ by taking asymptotic form of digamma function in the integrand. The Kondo temperature dependence on chemical potential and exchange coupling are obtained. Depending on the location of the impurities the Kondo contribution to resistivity is very different. For the type of magnetic
impurities which break the $C_{3v}$ symmetry at low chemical potential it shows power law temperature dependence as $1/(T-T_k)$. For the magnetic impurities
preserving the $C_{3v}$ symmetry of the lattice at low chemical potential it shows power law dependence as $1/(T^3-T_k^3)$. At even lower chemical potential when 
Fermi surface is close to the node, both cases show extra dependence on chemical potential as well as Kondo scale. Near the node a critical value of exchange coupling is needed for Kondo effect to be realized\cite{Carlos,DW}. The critical value $|J_c|$ is larger for symmetry breaking case. For large chemical potential $\mu$ both cases
show logarithmic dependence on temperature scaled by the Kondo temperature. With increasing $\mu$ the resistivity at a given temperature decreases for impurities breaking the honeycomb symmetry while the resistivity increases for the ones preserving the symmetry. The resistivity obtained with same set of parameters show that the dominant source of resistivity is from the impurities which break the $C_{3v}$ symmetry.

We also have computed the effect of charge fluctuation for impurity occupation $0\le n_{d,s}\ll 0.5$ by using mean field approach on the Anderson impurity model. 
The resistivity at a given temperature has similar dependence on the chemical potential as the case for spin fluctuations. Similar to the spin fluctuation case the dominant contribution to resistivity at the same sets of parameters comes from the impurities which break the honeycomb lattice symmetry.

By studying the resistivity v.s. temperature and comparing with experimental results in Ref.\onlinecite{Chen} from both spin and charge fluctuations in the symmetry breaking case we find that the Kondo effect fails to give the correct Kondo scale and unable to describe single parameter scaling for chemical potential nearby the node. For $\mu\sim 0$ the resistivity due to charge fluctuations give reasonable temperature dependence and the resistivity after rescaling also shows single parameter universal behavior. The same analysis yields large Kondo scale for $10^{-2}eV<\mu$ in the charge fluctuation case which is roughly the
same chemical potential at which we get non monotonic behavior of Kondo temperature in the spin fluctuation (Kondo) case. 

The failure of Kondo explanation nearby the node is consistent with the numerical RG results\cite{Ingersent} which find the Kondo effect near the node is suppressed for $r>1/2$ for systems having electronic density of state $\rho(\epsilon)\propto|\epsilon|^r$. By combining the low chemical potential results ($0\le\mu\le10^{-3}eV)$ from charge fluctuation with the large $10^{-2}eV<\mu$ results from Kondo effect obtain the Kondo scale consistent with the experimental results. 
For chemical potential in between these two cases the system should be in the mixed valence regime.
For gate voltage higher than $30V$ a weaker dependence of the impurity energy on the applied gate voltage as compared to the dependence at smaller chemical potentials will lead to a decrease in the Kondo temperature. Whether this effect or the effect of RKKY interactions is responsible for the observed non-monotonic behavior on gate voltage will be the subject of future studies.
\section*{Acknowledgment}
The authors wish to acknowledge Roland Kawakami and Shan-Wen Tsai for useful discussions. Vivek Aji and Sung-Po Chao's research is supported by University of California at Riverside under the initial complement.
\appendix
\section{Derivation for symmetry breaking case}\label{A}
\begin{eqnarray*}
R_{SB}(z)=\frac{4}{9\pi t^2}\int_{-\infty}^{\infty}\frac{|\epsilon|d\epsilon}{\epsilon-(z+\mu)}\tanh(\frac{\epsilon-\mu}{2 k_B T})\frac{\Lambda^2}{\epsilon^2+\Lambda^2}
\end{eqnarray*}
Define $y=\epsilon-\mu$ and use
$$\tanh(\frac{\beta y}{2})=\frac{1}{i\pi}(\psi(\frac{1}{2}+i\frac{\beta y}{2\pi})-\psi(\frac{1}{2}-i\frac{\beta y}{2\pi}))$$. As $\psi(\frac{1}{2}\pm i\frac{\beta y}{2\pi})$ has poles on upper/lower complex plane we may separate $R_{SB}(z)$ into two parts as $R_{SB}(z)=(I_1+I_2)4/(9\pi t^2)$ with $I_1$ and 
$I_2$ given by
\begin{eqnarray*}
I_1&=&\frac{1}{i\pi}\int_{-\infty}^{\infty}dy \frac{|y+\mu|}{y-z}\frac{\Lambda^2}{(y+\mu)^2+\Lambda^2}\psi(\frac{1}{2}+i\frac{\beta y}{2\pi})\\\nonumber
I_2&=&\frac{-1}{i\pi}\int_{-\infty}^{\infty}dy \frac{|y+\mu|}{y-z}\frac{\Lambda^2}{(y+\mu)^2+\Lambda^2}\psi(\frac{1}{2}-i\frac{\beta y}{2\pi})
\end{eqnarray*}
We may write $I_1=I_{11}+I_{12}$ with 
\begin{eqnarray*}
I_{11}&=&\frac{-1}{i\pi}\int_{-\infty}^{\mu}dy \frac{y+\mu}{y-z}\frac{\Lambda^2}{(y+\mu)^2+\Lambda^2}\psi(\frac{1}{2}+i\frac{\beta y}{2\pi})\\\nonumber
&=&-\frac{\Lambda}{2\pi}\int_{-\infty}^{-\mu}dy \frac{y+\mu}{y-z}(\frac{\psi(\frac{1}{2}+i\frac{\beta y}{2\pi})}{y+\mu+i\Lambda}-\frac{\psi(\frac{1}{2}+i\frac{\beta y}{2\pi})}{y+\mu-i\Lambda})\\\nonumber
&=&\Big\{\frac{\Lambda}{2\pi}i\pi\frac{-i\Lambda}{-i\Lambda-(z+\mu)}\psi(\frac{1}{2}-i\frac{\beta\mu}{2\pi}+\frac{\beta\Lambda}{2\pi})\\\nonumber
&+&\frac{\Lambda}{2\pi}\int_0^{-\Lambda+\delta}idx\frac{ix}{ix-(z+\mu)}\frac{1}{i(x+\Lambda)}\psi(\frac{1}{2}-i\frac{\beta\mu}{2\pi}-\frac{\beta x}{2\pi})\\\nonumber
&+&\frac{\Lambda}{2\pi}\int_{-\Lambda-\delta}^{-\infty}idx\frac{ix}{ix-(z+\mu)}\frac{1}{i(x+\Lambda)}\psi(\frac{1}{2}-i\frac{\beta\mu}{2\pi}-\frac{\beta x}{2\pi})\Big\}\\\nonumber
&+&\frac{\Lambda}{2\pi}\int_{-\infty}^{-\mu}dy \frac{y+\mu}{y-z}\frac{\psi(\frac{1}{2}+i\frac{\beta y}{2})}{y+\mu-i\Lambda}
\end{eqnarray*}
In the third and fourth lines of the above equation we replaced $y$ by $y=-\mu+i x$ and $\delta\rightarrow +0$. Similar computation gives $I_{12}$ as
\begin{eqnarray*}
I_{12}&=&\frac{\Lambda}{2\pi}\int_{-\mu}^{\infty}dy \frac{y+\mu}{y-z}(\frac{\psi(\frac{1}{2}+i\frac{\beta y}{2\pi})}{y+\mu+i\Lambda}-\frac{\psi(\frac{1}{2}+i\frac{\beta y}{2\pi})}{y+\mu-i\Lambda})\\
&=&(-\frac{\Lambda}{2\pi})\Big\{\int_{-\infty}^{-\Lambda-\delta}idx\frac{ix}{ix-(z+\mu)}\frac{\psi(\frac{1}{2}-i\frac{\beta\mu}{2\pi}-\frac{\beta x}{2\pi})}{i(x+\Lambda)}\\
&+&\int_{-\Lambda+\delta}^0 idx \frac{ix}{ix-(z+\mu)}\frac{\psi(\frac{1}{2}-i\frac{\beta\mu}{2\pi}-\frac{\beta x}{2\pi})}{i(x+\Lambda)}\Big\}\\
&-&i\pi\frac{\Lambda}{2\pi}\frac{i\Lambda}{(\mu+z)+i\Lambda}\psi(\frac{1}{2}-i\frac{\beta\mu}{2\pi}+\frac{\beta \Lambda}{2\pi})\\
&-&\frac{\Lambda}{2\pi}\int_{-\mu}^{\infty}dy\frac{y+\mu}{y-z}\frac{\psi(\frac{1}{2}+i\frac{\beta y}{2\pi})}{y+\mu-i\Lambda}
\end{eqnarray*}
Combining $I_{11}$ and $I_{12}$ we get
\begin{eqnarray*}
I_1
&=&\frac{\Lambda}{2\pi}\int_{\bar{C_1}} d\bar{z}\frac{\bar{z}+\mu}{\bar{z}-z}(\frac{\psi(\frac{1}{2}+i\frac{\beta\bar{z}}{2\pi})}{\bar{z}+\mu-i\Lambda})\times 2\\
&-&\frac{\Lambda}{i\pi}\Big(\int_{-\infty}^{-\Lambda-\delta}+\int_{-\Lambda+\delta}^{0}\Big) dx\frac{x \psi(\frac{1}{2}-i\frac{\beta\mu}{2\pi}-\frac{\beta x}{2\pi}) }{(x+\Lambda)(z+\mu-ix)}
\end{eqnarray*}
Here $\bar{C_1}$ denotes the integration path taken from $\bar{z}=-\mu-i\infty$ to $\bar{z}=-\mu$ along imaginary axis. We may also write $I_2=I_{21}+I_{22}$ in different region as
\begin{eqnarray*}
I_{21}&=&\frac{1}{i\pi}\int_{-\infty}^{-\mu}dy \frac{y+\mu}{y-z}\frac{\Lambda^2}{(y+\mu)^2+\Lambda^2}\psi(\frac{1}{2}-i\frac{\beta y}{2\pi})\\
&=&\frac{\Lambda}{2\pi}\int_{-\infty}^{-\mu}dy \frac{y+\mu}{y-z}(\frac{\psi(\frac{1}{2}-i\frac{\beta y}{2\pi})}{y+\mu+i\Lambda}-\frac{\psi(\frac{1}{2}-i\frac{\beta y}{2\pi})}{y+\mu-i\Lambda})\\
&=&\frac{-\Lambda}{2\pi}\int_{\bar{C_2}}d\bar{z}\frac{\bar{z}+\mu}{\bar{z}-z}\frac{\psi(\frac{1}{2}-i\frac{\beta\bar{z}}{2\pi})}{\bar{z}+\mu+i\Lambda}\\
&+&\frac{\Lambda}{2\pi}\Big(\int_{0}^{\Lambda-\delta}+\int_{\Lambda+\delta}^{\infty}\Big)  dx\frac{x \psi(\frac{1}{2}+i\frac{\beta\mu}{2\pi}+\frac{\beta x}{2\pi})}{x+i(\mu+z)(x-\Lambda)}\\
&-&i\pi\frac{\Lambda}{2\pi}\frac{i\Lambda}{i\Lambda-(\mu+z)}\psi(\frac{1}{2}+i\frac{\beta\mu}{2\pi}+\frac{\beta\Lambda}{2\pi})
\end{eqnarray*}
Here $\bar{C_2}$ denotes the integration path taken from $\bar{z}=-\mu$ to $\bar{z}=-\mu+i\infty$ along imaginary axis. $I_{22}$ is expressed as
\begin{eqnarray*}
I_{22}&=&\frac{-1}{i\pi}\int_{-\mu}^{\infty}dy \frac{y+\mu}{y-z}\frac{\Lambda^2}{(y+\mu)^2+\Lambda^2}\psi(\frac{1}{2}-i\frac{\beta y}{2\pi})\\
&=&\frac{\Lambda}{2\pi}\int_{-\mu}^{\infty}dy \frac{y+\mu}{y-z}(\frac{\psi(\frac{1}{2}-i\frac{\beta y}{2\pi})}{y+\mu-i\Lambda}-\frac{\psi(\frac{1}{2}-i\frac{\beta y}{2\pi})}{y+\mu+i\Lambda})\\
&=&-\frac{\Lambda}{2\pi}2\pi i\frac{z+\mu}{z+\mu+i\Lambda}\psi(\frac{1}{2}-i\frac{\beta z}{2\pi})\\
&-&\frac{\Lambda}{2\pi}\int_{\bar{C_2}}d\bar{z}\frac{\bar{z}+\mu}{\bar{z}-z}\frac{\psi(\frac{1}{2}-i\frac{\beta\bar{z}}{2\pi})}{\bar{z}+\mu+i\Lambda}\\
&-&\frac{\Lambda}{2\pi}\Big(\int_{\infty}^{\Lambda+\delta}+\int_{\Lambda-\delta}^0\Big) dx \frac{x\psi(\frac{1}{2}+i\frac{\beta\mu}{2\pi}+\frac{\beta x}{2\pi})}{(x+i(\mu+z))(x-\Lambda)}\\
&+&\frac{\Lambda}{2\pi}i\pi\frac{i\Lambda}{i\Lambda-(\mu+z)}\psi(\frac{1}{2}+i\frac{\beta\mu}{2\pi}+\frac{\beta\Lambda}{2\pi})\\
&+&\frac{\Lambda}{2\pi}2\pi i\frac{z+\mu}{z+\mu-i\Lambda}\psi(\frac{1}{2}-i\frac{\beta z}{2\pi})
\end{eqnarray*}
The sum of $I_1$ and $I_2$ is then given by
\begin{eqnarray*}
&&R_{SB}(z)=\frac{4}{9\pi t^2}(I_1+I_2)\\
&&=\frac{4}{9\pi t^2}\Big\{\frac{\Lambda}{\pi}\Big(\int_{\bar{C_1}}d\bar{z}\frac{\bar{z}+\mu}{\bar{z}-z}\frac{\psi(\frac{1}{2}+i\frac{\beta\bar{z}}{2\pi})}{\bar{z}+\mu-i\Lambda}\\&&
-\int_{\bar{C_2}}d\bar{z}\frac{\bar{z}+\mu}{\bar{z}-z}\frac{\psi(\frac{1}{2}-i\frac{\beta\bar{z}}{2\pi})}{\bar{z}+\mu+i\Lambda}\Big)\\
&&-\frac{\Lambda}{\pi i}\Big(\int_{-\infty}^{-\Lambda-\delta}+\int_{-\Lambda+\delta}^0\Big) dx\frac{x\psi(\frac{1}{2}-i\frac{\beta\mu}{2\pi}-\frac{\beta x}{2\pi})}
{(z+\mu-ix)(x+\Lambda)}\\
&&+\frac{\Lambda}{\pi i}\Big(\int_{0}^{\Lambda-\delta}+\int_{\Lambda+\delta}^\infty\Big) dx\frac{x\psi(\frac{1}{2}+i\frac{\beta\mu}{2\pi}+\frac{\beta x}{2\pi})}
{(z+\mu-ix)(x-\Lambda)}\\
&&+i\Lambda\psi(\frac{1}{2}-i\frac{\beta z}{2\pi})(\frac{z+\mu}{z+\mu-i\Lambda}-\frac{z+\mu}{z+\mu+i\Lambda})\Big\}
\end{eqnarray*}
Rewrite $\bar{z}=-\mu+ix$ in the expression of $R_{SB}(z)$ along the $\bar{C_1}$ and $\bar{C_2}$ paths we get
\begin{eqnarray*}
R_{SB}(z)&=&\frac{4}{9\pi t^2}\Big\{\frac{-\Lambda}{i\pi}\Big(\int_0^{\infty}dx \frac{x\psi(\frac{1}{2}+i\frac{\beta\mu}{2\pi}+\frac{\beta x}{2\pi})}{(\mu+z-ix)(x+\Lambda)}\\&-&\int_0^{\infty}dx \frac{x\psi(\frac{1}{2}-i\frac{\beta\mu}{2\pi}+\frac{\beta x}{2\pi})}{(\mu+z+ix)(x+\Lambda)}\Big)\\
&+&\frac{\Lambda}{i\pi}\Big(\int_0^{\Lambda-\delta}+\int_{\Lambda+\delta}^{\infty}\Big)dx\Big[\frac{x\psi(\frac{1}{2}+i\frac{\beta\mu}{2\pi}+\frac{\beta x}{2\pi})}{(\mu+z-ix)(x+\Lambda)}\\&-&\frac{x\psi(\frac{1}{2}-i\frac{\beta\mu}{2\pi}+\frac{\beta x}{2\pi})}{(\mu+z+ix)(x+\Lambda)}\Big]\\
&+&\psi(\frac{1}{2}-i\frac{\beta z}{2\pi})\Big[\frac{-2\Lambda^2(z+\mu)}{(z+\mu)^2+\Lambda^2}\Big]\Big\}
\end{eqnarray*}
By defining $F(x,\mu,z)$ as 
$$F(x,\mu,z)=\frac{\psi(\frac{1}{2}+i\frac{\beta\mu}{2\pi}+\frac{\beta x}{2\pi})}{x+i(\mu+z)}+
\frac{\psi(\frac{1}{2}-i\frac{\beta\mu}{2\pi}+\frac{\beta x}{2\pi})}{x-i(\mu+z)}$$
we may simplify above expression as
\begin{eqnarray}\nonumber
R_{SB}(z)&=&\frac{4}{9\pi t^2}\Big\{\frac{\Lambda}{\pi}\Big(\mathrm{P}\int_0^{\infty} dx \frac{F(x,\mu,z)x}{x-\Lambda}\\\nonumber&-&\int_0^{\infty} dx \frac{F(x,\mu,z)x}{x+\Lambda}\Big)\\&-&\psi(\frac{1}{2}-i\frac{\beta z}{2\pi})\frac{2\Lambda^2(z+\mu)}{(z+\mu)^2+\Lambda^2}\Big\}
\end{eqnarray}
\section{Derivation for symmetry preserving case}\label{B}
Consider integrals of the form:
\begin{eqnarray*}
R_{SP}(z)=\frac{4}{9\pi t^4}\int_{-\infty}^{\infty}\frac{|\epsilon|^3 d\epsilon}{\epsilon-(z+\mu)}\tanh(\frac{\epsilon-\mu}{2k_B T})\frac{\Lambda^4}{\epsilon^4+\Lambda^4}
\end{eqnarray*}
let $y=\epsilon-\mu$ we get
\begin{eqnarray*}
R_{SP}(z)=\frac{4}{9\pi t^4}\Big\{\frac{1}{i\pi}\int_{-\infty}^{\infty}dy \frac{|y+\mu|^3}{y-z}\frac{\Lambda^4}{(y+\mu)^4+\Lambda^4}\\\times[\psi(\frac{1}{2}+\frac{i\beta y}{2\pi})-
\psi(\frac{1}{2}-\frac{i\beta y}{2\pi})]\Big\}=\frac{4}{9\pi t^4}(\bar{I}_1+\bar{I}_2)
\end{eqnarray*}
We take the integration regions into two parts by writing 
$\bar{I}_1=\bar{I}_{11}+\bar{I}_{12}$
with 
\begin{eqnarray*}
&&\bar{I}_{11}=\frac{-1}{i\pi}\int_{-\infty}^{-\mu}dy \frac{(y+\mu)^3}{y-z}\frac{\Lambda^4}{(y+\mu)^4+\Lambda^4}\psi(\frac{1}{2}+\frac{i\beta y}{2\pi})\\
&&=\frac{-1}{i\pi}\Big[-2\pi i\frac{(\Lambda e^{-\frac{3}{4}\pi i})^3}{\Lambda e^{-\frac{3}{4}\pi i}-(z+\mu)}\times\\&&\frac{\Lambda^4\psi(\frac{1}{2}-i\frac{\beta\mu}{2\pi}+i\frac{\beta\Lambda e^{-\frac{3}{4}\pi i}}{2\pi})}{\Lambda^3(e^{-\frac{3}{4}\pi i}-e^{\frac{1}{4}\pi i})(e^{-\frac{3}{4}\pi i}-e^{\frac{-1}{4}\pi i})(e^{-\frac{3}{4}\pi i}-e^{\frac{3}{4}\pi i})}\\&&-\int_0^{-\infty}dx\frac{x^3}{ix-(z+\mu)}\Big[\frac{\Lambda^4\psi(\frac{1}{2}-i\frac{\beta\mu}{2\pi}-\frac{\beta x}{2\pi}}{x^4+\Lambda^4}\Big]\Big]
\end{eqnarray*}
\begin{eqnarray*}
&&\bar{I}_{12}=\frac{1}{i\pi}\int_{-\mu}^{\infty}dy \frac{(y+\mu)^3}{y-z}\frac{\Lambda^4}{(y+\mu)^4+\Lambda^4}\psi(\frac{1}{2}+\frac{i\beta y}{2\pi})\\
&&=\frac{1}{i\pi}\Big[-2\pi i\frac{(\Lambda e^{-\frac{1}{4}\pi i})^3}{\Lambda e^{-\frac{1}{4}\pi i}-(z+\mu)}\times\\&&\frac{\Lambda^4\psi(\frac{1}{2}-i\frac{\beta\mu}{2\pi}+i\frac{\beta\Lambda e^{-\frac{1}{4}\pi i}}{2\pi})}{\Lambda^3(e^{-\frac{1}{4}\pi i}-e^{\frac{1}{4}\pi i})(e^{-\frac{1}{4}\pi i}-e^{\frac{3}{4}\pi i})(e^{-\frac{1}{4}\pi i}-e^{\frac{-3}{4}\pi i})}\\&&-\int_{-\infty}^0 dx\frac{x^3}{ix-(z+\mu)}\Big[\frac{\Lambda^4\psi(\frac{1}{2}-i\frac{\beta\mu}{2\pi}-\frac{\beta x}{2\pi}}{x^4+\Lambda^4}\Big]\Big]
\end{eqnarray*}
Thus 
\begin{eqnarray*}
&&\bar{I}_1=\bar{I}_{11}+\bar{I}_{12}\\
&&=\frac{\Lambda^4\psi(\frac{1}{2}-i\frac{\beta\mu}{2\pi}+\frac{i\beta\Lambda e^{-\frac{3}{4}\pi i}}{2\pi})}{2(\Lambda e^{-\frac{3}{4}\pi i}-(z+\mu))}
-\frac{\Lambda^4\psi(\frac{1}{2}-i\frac{\beta\mu}{2\pi}+\frac{i\beta\Lambda e^{-\frac{\pi i}{4}}}{2\pi})}{2(\Lambda e^{-\frac{\pi i}{4}}-(z+\mu))}\\
&&+\frac{2}{i\pi}\int_0^{-\infty}dx\frac{x^3}{ix-(z+\mu)}\frac{\Lambda^4\psi(\frac{1}{2}-i\frac{\beta\mu}{2\pi}-\frac{\beta x}{2\pi})}{x^4+\Lambda^4}
\end{eqnarray*}
Similarly we can write $\bar{I}_2=\bar{I}_{21}+\bar{I}_{22}$
with 
\begin{eqnarray*}
&&\bar{I}_{21}=\frac{1}{i\pi}\int_{-\infty}^{-\mu}dy \frac{(y+\mu)^3}{y-z}\frac{\Lambda^4}{(y+\mu)^4+\Lambda^4}\psi(\frac{1}{2}-\frac{i\beta y}{2\pi})\\
&&=\frac{1}{i\pi}\Big\{2\pi i\frac{(\Lambda e^{\frac{3\pi i}{4}})^3}{\Lambda e^{\frac{3\pi i}{4}}-(z+\mu)}\times\\
&&\frac{\Lambda\psi(\frac{1}{2}+i\frac{\beta\mu}{2\pi}-i\frac{\beta\Lambda}{2\pi} e^{\frac{3\pi i}{4}})}{(e^{\frac{3\pi i}{4}}-e^{\frac{\pi i}{4}})(e^{\frac{3\pi i}{4}}-e^{\frac{-\pi i}{4}})(e^{\frac{3\pi i}{4}}-e^{\frac{-3\pi i}{4}})}\\
&&-\int_0^{\infty}dx \frac{x^3}{i x-(z+\mu)}\Big[\frac{\Lambda^4\psi(\frac{1}{2}+i\frac{\beta\mu}{2\pi}+\frac{\beta x}{2\pi})}{x^4+\Lambda^4}\Big]\Big\}
\end{eqnarray*}
and 
\begin{eqnarray*}
&&\bar{I}_{22}=\frac{-1}{i\pi}\int_{-\mu}^{\infty}dy \frac{(y+\mu)^3}{y-z}\frac{\Lambda^4}{(y+\mu)^4+\Lambda^4}\psi(\frac{1}{2}-\frac{i\beta y}{2\pi})\\
&&=\frac{1}{i\pi}\Big\{2\pi i\frac{(\Lambda e^{\frac{\pi i}{4}})^3}{\Lambda e^{\frac{\pi i}{4}}-(z+\mu)}\times\\
&&\frac{\Lambda\psi(\frac{1}{2}+i\frac{\beta\mu}{2\pi}-i\frac{\beta\Lambda}{2\pi} e^{\frac{\pi i}{4}})}{(e^{\frac{\pi i}{4}}-e^{\frac{-\pi i}{4}})(e^{\frac{\pi i}{4}}-e^{\frac{3\pi i}{4}})(e^{\frac{\pi i}{4}}-e^{\frac{-3\pi i}{4}})}\\
&&+2\pi i\frac{(\mu+z)^3\Lambda^4}{(\mu+z)^4+\Lambda^4}\psi(\frac{1}{2}-i\frac{\beta z}{2\pi})\\
&&-\int_{\infty}^0 dx\frac{x^3}{ix-(z+\mu)}\Big[\frac{\Lambda^4\psi(\frac{1}{2}+i\frac{\beta\mu}{2\pi}+\frac{\beta x}{2\pi})}{x^4+\Lambda^4}\Big]\Big\}
\end{eqnarray*}
Thus
\begin{eqnarray*}
&&\bar{I}_2=\bar{I}_{21}+\bar{I}_{22}\\
&&=\frac{\Lambda^4\psi(\frac{1}{2}+i\frac{\beta\mu}{2\pi}-i\frac{\beta\Lambda}{2\pi} e^{\frac{3\pi i}{4}})}{2(\Lambda e^{\frac{3\pi i}{4}}-(z+\mu))}
-\frac{\Lambda^4\psi(\frac{1}{2}+i\frac{\beta\mu}{2\pi}-i\frac{\beta\Lambda}{2\pi} e^{\frac{\pi i}{4}})}{2(\Lambda e^{\frac{\pi i}{4}}-(z+\mu))}\\
&&-2\frac{(z+\mu)^3\Lambda^4}{(z+\mu)^4+\Lambda^4}\psi(\frac{1}{2}-i\frac{\beta z}{2\pi})\\
&&-\frac{2}{i\pi}\int_0^{\infty}dx\frac{x^3}{ix-(z+\mu)}\Big[\frac{\Lambda^4\psi(\frac{1}{2}+i\frac{\beta\mu}{2\pi}+\frac{\beta x}{2\pi})}{x^4+\Lambda^4}\Big]
\end{eqnarray*}
We combine results of $\bar{I}_1$ and $\bar{I}_2$ to get
\begin{eqnarray*}
&&R_{SP}(z)=\frac{4}{9\pi t^4}(\bar{I}_1+\bar{I}_2)\\
&&=\frac{16}{81\pi t^4}\Big\{\frac{2\Lambda^4}{\pi}\int_0^{\infty}dx \frac{x^3}{x^4+\Lambda^4}F(x,\mu,z)\\&&-2\frac{(z+\mu)^3\Lambda^4}{(z+\mu)^4+\Lambda^4}\psi(\frac{1}{2}-i\frac{\beta z}{2\pi})\\
&&+\Re\Big[\frac{\Lambda^4\psi(\frac{1}{2}+i\frac{\beta\mu}{2\pi}-i\frac{\beta\Lambda}{2\pi} e^{\frac{3\pi i}{4}})}{\Lambda e^{\frac{3\pi i}{4}}-(z+\mu)}
-\frac{\Lambda^4\psi(\frac{1}{2}+i\frac{\beta\mu}{2\pi}-i\frac{\beta\Lambda}{2\pi} e^{\frac{\pi i}{4}})}{\Lambda e^{\frac{\pi i}{4}}-(z+\mu)}\Big]\Big\}
\end{eqnarray*}

\end{document}